\documentclass[Journal=jctcce,manuscript=article]{achemso}

\usepackage[version=3]{mhchem} 
\usepackage{bm}
\usepackage{xcolor}
\usepackage{dsfont}
\usepackage[left=2cm,right=2cm,top=3cm,bottom=2cm]{pdflscape}
\usepackage{graphicx}
\usepackage{subcaption}

\author{Joseph P. Heindel}
\affiliation{Kenneth S. Pitzer Theory Center and Department of Chemistry, University of California, Berkeley, California 94720, United States}
\alsoaffiliation{Chemical Sciences Division, Lawrence Berkeley National Laboratory, Berkeley, California 94720, United States}
\author{Selim Sami}
\affiliation{Kenneth S. Pitzer Theory Center and Department of Chemistry, University of California, Berkeley, California 94720, United States}
\author{Teresa Head-Gordon}
\email{heindelj@lbl.gov, thg@berkeley.edu}
\affiliation{Kenneth S. Pitzer Theory Center and Department of Chemistry, University of California, Berkeley, California 94720, United States}
\alsoaffiliation{Chemical Sciences Division, Lawrence Berkeley National Laboratory, Berkeley, California 94720, United States}
\alsoaffiliation{Departments of Bioengineering and Chemical and Biomolecular Engineering, University of California, Berkeley, California 94720, United States}

\title[An \textsf{achemso} demo]
  {Completely Multipolar Model as a General Framework for Many-Body Interactions as Illustrated for Water}


\begin{document}

\begin{tocentry}
\centering
\includegraphics[width=1.05\textwidth]{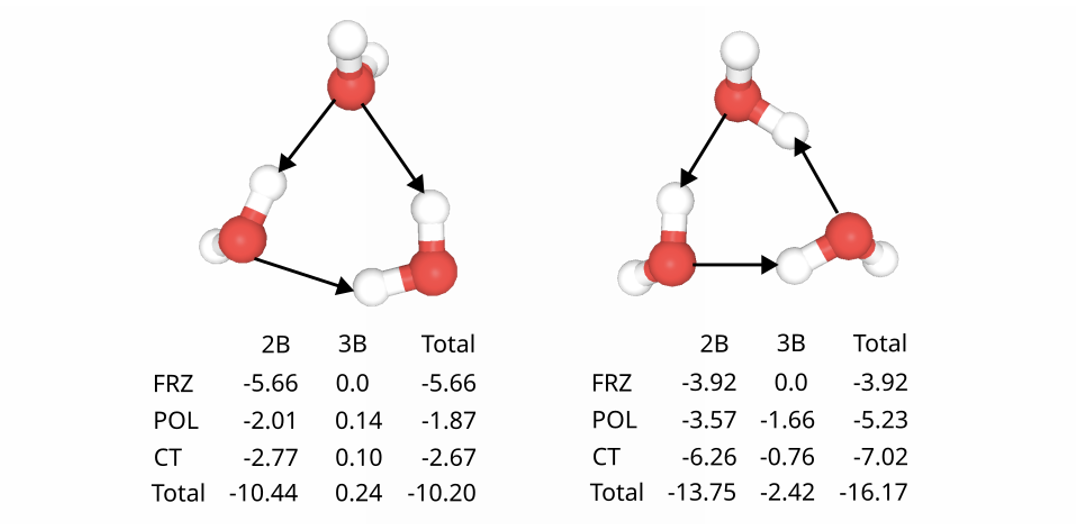}
\end{tocentry}

\begin{abstract}
\noindent
We introduce a general framework for many-body force field models, the Completely Multipolar Model (CMM), that utilizes multipolar electrical moments modulated by exponential decay of electron density as a common functional form for all piecewise terms of an energy decomposition analysis of intermolecular interactions. With this common functional form the CMM model establishes well-formulated damped tensors that reach the correct asymptotes at both long- and short-range while formally ensuring no short-range catastrophes. The CMM describes the separable EDA terms of dispersion, exchange polarization, and Pauli repulsion with short-ranged anisotropy, polarization as intramolecular charge fluctuations and induced dipoles, while charge transfer describes explicit movement of charge between molecules, and naturally describes many-body charge transfer by coupling into the polarization equations. We also utilize a new one-body potential that accounts for intramolecular polarization by including an electric field-dependent correction to the Morse potential to ensure that the CMM reproduces all physically relevant monomer properties including the dipole moment, molecular polarizability, and dipole and polarizability derivatives. The quality of the CMM is illustrated through agreement of individual terms of the EDA and excellent extrapolation to energies and geometries of an extensive validation set of water cluster data. 
\end{abstract}

\section*{Introduction}
Force fields (FFs) are approximations to the quantum mechanical (QM) potential energy surface, in which the model design goal is to predict structure, dynamics, and thermodynamics of any molecular system. Although pairwise additive FFs remain popular due to their computational efficiency, they are limited by their inability to describe the inherent many-body contributions of the QM energy. This greatly reduces their accuracy and transferability in property prediction when traversing the phase diagram of a homogeneous system such as water, and extensions to  heterogenous systems. 

Recently, there has been a paradigm shift in non-reactive many-body FF development by combining energy decomposition analysis (EDA)\cite{Szalewicz:2012:SAPT,Patkowski:2020:SAPT,Mao:2021:EDA-review} with the many-body expansion (MBE)\cite{demerdash2017assessing,heindel2020many,heindel2023many} to better control the accuracy and transferability of advanced FF models. More specifically, the EDA scheme based on absolutely localized molecular orbitals (ALMO-EDA) decomposes the total QM energy into physically motivated contributions\cite{khaliullin2007,horn2016probing}
\begin{equation}
E_{\rm int} = E_{\rm Pauli} + E_{\rm Elec,CP} + E_{\rm Disp} + E_{\rm Pol,Exch\_Pol} +  E_{\rm CT}
 \label{eq:eda}
\end{equation}
such as pairwise additive terms due to Pauli repulsion ($E_{\rm Pauli}$) and electrostatics with charge penetration ($E_{\rm elec,CP}$), as well as many-body contributions that arise from dispersion ($E_{\rm disp}$), polarization and exchange polarization ($ E_{\rm Pol,Exch{\_}Pol}$), and charge transfer ($E_{\rm CT}$). In turn the MBE of the non-bonded energy\cite{heindel2020many,heindel2021many,herman2021many} allows one to gain information as to how much non-additivity is present in the decomposed QM energy (and forces\cite{aldossary2023force}). For example, our recent MB-UCB force field for water\cite{das2019development} and extensions to monovalent and divalent alkali metal and halide ions\cite{das2022development} utilized a systematic buildup of modeling 2- and 3-body  molecular interactions formulated from the ALMO-EDA energy decomposition scheme.\cite{mao2021intermolecular} By reproducing the ALMO-EDA term-by-term, we can ensure that the force field will be transferable across different state points as we showed for water\cite{das2019development}, and to heterogeneous atomic ion-water systems\cite{das2022development}.

While many-body potentials for water have reached high accuracy, FFs such as MB-Pol\cite{babin2013development,babin2014development}, MASTIFF\cite{van2016beyond,van2018new}, AMOEBA+\cite{Liu2019}, HIPPO\cite{rackers2021polarizable}, MB-UCB\cite{das2019development,das2022development}, q-AQUA\cite{yu2022q}, and others\cite{Fanourgakis2006, Fanourgakis2008,Davie2016,Konovalov2021,Naseem-Khan2022} differ in their use of EDA, MBE, and most critically, their functional forms for the decomposed energy and forces. While MB-Pol and q-AQUA rely on the MBE, they do not use EDA and largely relegate modeling most or all short-ranged interactions as fits to high-order polynomials, which limits their transferability to other systems. Models like AMOEBA+, MB-UCB, and CHARMM models have different approaches to including polarization\cite{Demerdash2014,Demerdash2018}, such as Drude oscillators\cite{drude1902} and related core-shell models\cite{Leven2019}, fluctuating charges\cite{rick1994} and related bond capacity models\cite{poier2019describing,poier2019molecular,jensen2023unifying}, and induced dipoles\cite{applequist1985multipole,thole1981molecular}. There have also been unified approaches allowing for both charge rearrangements and induced dipoles\cite{stern2001combined,giovannini2019polarizable} in order to also capture the charge transfer interaction.\cite{rick1994,Naseem-Khan2022} 
But with the advent of variational EDA techniques such as ALMO-EDA that isolate charge transfer\cite{khaliullin2007,horn2016probing,Mao:2021:EDA-review}, it is now clear that the charge transfer energy scales exponentially and hence cannot be modelled by fluctuating charges or induced dipole models alone. 

Other advanced force fields rely on the density overlap hypothesis\cite{kim1981dependence,wheatley1990overlap,gavezzotti2002calculation,van2016beyond,van2018new}
which states that the short-range contributions to intermolecular interactions is proportional to the electron density overlap. The density overlap hypothesis has been advanced by Misquitta and others\cite{misquitta2014distributed,misquitta2018isa} based on iterated stockholder atoms which can be used to define Slater-like densities for atoms in molecules, as well as by van Vleet \textit{et al.} in their MASTIFF force field.\cite{van2016beyond,van2018new} Rackers \textit{et al.} utilize a similar idea in the HIPPO model\cite{rackers2021polarizable} but rather than relying on density overlap, they treat the Slater function as an orbital and are able to derive models of Pauli repulsion, charge penetration, dispersion, along with well motivated damping functions. 

This work takes a step forward in a systematic and unified construction of a general functional form for all pairwise and many-body contributions to energies and forces, as illustrated for water. 
We describe all piecewise terms of ALMO-EDA in terms of multipolar electrical moments, modulated by exponential decay of electron density, but acting at different spatial ranges as controlled by newly defined damped tensor interactions as an alternative to density overlap formulations. The resulting Completely Multipolar Model (CMM) relies on differences in undamped and damped tensors for Pauli repulsion and charge transfer terms, and introduces a correct polynomial order for the damped tensors for polarization that is necessary to avoid all short-range catastrophes. Additionally the CMM charge transfer model describes explicit forward and backward transfer of charge between molecules, and naturally describes many-body charge transfer by coupling into the polarization equations. Finally, we eliminate the need to treat intramolecular and intermolecular polarization separately through our recently reported one-body potential\cite{Sami2024}, further modified by a field-dependent correction to the Morse potential, thereby further improving the accuracy of electrostatic and polarization forces and ability to model spectroscopy. The CMM model for water shows excellent accuracy against EDA data and extrapolates well to an independent validation set of water clusters energies and geometries, and reproduces the structure-frequency correspondence expected for hydrogen-bonded \ce{O-H} stretches.\cite{boyer2019beyond} The CMM approach also serves as a validation of the ALMO-EDA that inherently captures a separation of charge transfer and exchange and many-body polarization, but in a regime where covalency is minimized so that interactions remain truly intermolecular and thus separable. 

\section*{Theory}
\noindent
Over the full range of interaction distances $r$, there are two limiting regimes which all non-reactive FF functional forms should obey as intermolecular interactions become either long- or short-ranged. The long-range limit is the standard asymptotics as $r\gg r_{eq}$, where $r_{eq}$ is the equilibrium intermolecular distance between two atoms, while the short-ranged limit is sometimes referred to as the united atom limit since it describes the case where the nuclei overlap. The united atom limit can be seen to be relevant to non-reactive force fields by considering a bond breaking process during which the involved atoms are highly polarized compared to either reference bonding state. Many polarizable models do not appropriately damp such strong interactions such that they can formally diverge. We will show that the CMM framework can guarantee singularity-free polarization which greatly improves stability and accuracy. These observations are especially important for a future reactive FF in order to smoothly change intermolecular interactions into intramolecular interactions (although not pursued here).

Table \ref{tab:asymptote} shows that the long-ranged and short-ranged interactions differ for each term in an EDA, and while most advanced FFs respect the long-range asymptotes, it is the treatment of short-ranged interactions that requires special consideration. Table \ref{tab:asymptote} emphasizes that some terms decay to a constant whereas others decay exponentially, and the signs of Pauli, CT, and Exch-Pol differ in the short range, all of which will be necessary to consider in the functional form. We note that the Exch-Pol energy is part of $E_{pol}$ in Eq. \ref{eq:eda} as it describes the relief of Pauli repulsion due to polarization.
\begin{table}[H]
    \centering
    \begin{tabular}{ccccccc}
         & Elec & Disp & CT & Pauli & Pol & Exch-Pol \\
        $r \ll r_{eq}$ & $1/r$ & $1/r^6$ & 0       & 0 & $1/r^n$ & 0 \\
        $r \gg r_{eq}$     & C     & C & $-e^{-r}$ & $e^{-r}$ & C & $-e^{-r}$ \\
    \end{tabular}
    \caption{\textit{Appropriate short and long-range limits of all considered interactions of an EDA.} Elec, Disp, Pauli, CT, Pol, and Exch-Pol (the very short-range between Pauli and polarization) correspond to Eq. (1). Note that $r_{eq}$ refers to an equilibrium intermolecular distance, not the equilibrium distance of a bond. The long-range scaling of polarization depends on the molecules, and we use the appropriate scaling of $n=6$ for water in this paper.}
    \label{tab:asymptote}
\end{table}


It is well-known that as interactions become increasingly short-ranged, multipole expansions do not converge.\cite{stone2013theory} This divergence arises in the region of density overlap and represents a breakdown of the point multipole representation of electrostatics. The simplest way to alleviate this problem is to give the multipoles a finite width to describe the anisotropic charge density.\cite{giese2008contracted,misquitta2008first,van2016beyond,van2018new,rackers2021polarizable} In this work we choose a Slater-like charge density 
\begin{equation}
  \rho(r)=\frac{Qb^3}{8\pi}e^{-br}+Z\delta(r)
  \label{eq:slater}
\end{equation}
where $Q$ is the charge associated with the model electron density, $Z$ is the effective nuclear charge of the atom, and $b$ defines the width of the Slater function. 

Rackers and Ponder have derived the relevant one-center and two-center damping functions for the Slater density.\cite{rackers2021polarizable} The general form of the damping and overlap functions for the one-center and two-center cases are,
\begin{subequations}
    \begin{equation}
        f^{damp}_i(r_{ij})=1-\left(1+\frac{1}{2}(b_ir_{ij})\right)e^{-b_ir_{ij}}
        \label{eq:f_damp}
    \end{equation}
    \begin{equation}
        f^{overlap}_{ij}(r_{ij})=1-\left(1+\frac{11}{16}(b_{ij}r_{ij})+\frac{3}{16}(b_{ij}r_{ij})^2+\frac{1}{48}(b_{ij}r_{ij})^3\right)e^{-b_{ij}r_{ij}}
        \label{eq:f_overlap}
    \end{equation}
\end{subequations}
\noindent
Note that $b_i$ is the range-parameter controlling the width of the electron density of atom $i$. For reasons we will present below, this width should not necessarily be the same for different classes of interactions. Furthermore, Eq. \ref{eq:f_overlap} is actually the solution to the integral when $b_i=b_j$ since the solution for inequivalent atoms has a more complicated form.\cite{rackers2021polarizable} For similar values of $b$, however, the integral is approximated well by the combination rule $b_{ij}=\sqrt{b_ib_j}$ which is the approach used here. The explicit form of the damping functions generated by the gradients were first reported elsewhere\cite{rackers2021polarizable} but are reproduced in the Supplementary Information for completeness.

We can now expand the Coulomb potential in gradients of the damped potential which result in the Coulomb interaction tensors for the one-center case, $\bm{T}_{ij}^{damp}$,
\begin{equation}
    \bm{T}_{ij}^{damp}=
    \begin{bmatrix}
        1 & \nabla & \nabla^2
    \end{bmatrix}\left(\frac{f_{ij}^{damp}(r_{ij})}{r_{ij}}\right)
    \label{eq:tensor_1}
\end{equation}
\noindent
Similarly, for the two-center case we get the overlap interaction tensors, $\bm{T}_{ij}^{overlap}$,
\begin{equation}
    \bm{T}_{ij}^{overlap}=
    \begin{bmatrix}
        1 & \nabla & \nabla^2 \\
        \nabla & \nabla^2 & \nabla^3 \\
        \nabla^2 & \nabla^3 & \nabla^4 \\
    \end{bmatrix}\left(\frac{f_{ij}^{overlap}(r_{ij})}{r_{ij}}\right)
    \label{eq:tensor_2}
\end{equation}
\noindent
In our approach, Eqs.  \ref{eq:tensor_1} and \ref{eq:tensor_2} generate the damped interaction tensors used for evaluating permanent electrostatics and polarization, but we will also see the overlap tensors make an appearance in the Pauli, dispersion, charge transfer, and exchange polarization terms. These damped interaction tensors have the same form as undamped Coulomb interaction tensors except that various entries in the tensor are scaled. 

More specifically, each term of the EDA will be determined by the scaling of a multipolar electrostatic interaction tensor which, at all orders, has the general form $P_n(br_{ij})e^{-br_{ij}}/r_{ij}$, where $P_n(br_{ij})$ is an $n$th order polynomial in powers of $br_{ij}$. Both $P_n(br_{ij})$ and the denominator $r_{ij}$ increase in order at each multipole rank so that the total scaling of each rank is the same; we will show that this is critical for avoiding polarization catastrophes, which is not the case for other force fields such as AMOEBA+\cite{Liu2019}, HIPPO\cite{rackers2021polarizable}, and MB-UCB\cite{das2019development,das2022development}. Although we will describe dispersion by a regular damped interaction tensor based on the Tang-Toennies form\cite{tang1992damping}, we show that it is directly related to Eq. \ref{eq:tensor_2}. For Pauli, CT, and Exch-Pol, a surprisingly simple idea is to formulate the difference between the damped multipolar interaction tensors and the undamped tensor as follows,
\begin{subequations}
\begin{equation}
    \mathbf{T}^{sr,+}_{ij}=(\mathbf{T}_{ij}-\mathbf{T}_{ij}^{overlap})
    \label{eq:mult_sr_plus}
\end{equation}
\begin{equation}
    \mathbf{T}^{sr,-}_{ij}=(\mathbf{T}_{ij}^{overlap}-\mathbf{T}_{ij})
    \label{eq:mult_sr_minus}
\end{equation}
\end{subequations}
that will maintain the short-ranged character of these terms. Notice that $\mathbf{T}^{sr,+}_{ij}$ is strictly positive for two charges of the same sign and will be used for Pauli repulsion, while $\mathbf{T}^{sr,-}_{ij}$ is strictly negative for two charges of the same sign and  will be used for CT and Exch-Pol that are strictly attractive. Finally, we allow each short-range term of the EDA in the CMM model to have a different value of $b$, a physical motivation related to the observation that coupling between exchange and each type of interaction occurs at different spatial ranges. \cite{tang1992damping}

Summarizing, CMM will use these general damping tensors to model each of the terms in Eq. \ref{eq:eda} for water using ALMO-EDA to separate the total non-bonded interaction energy into individual contributions. ALMO-EDA is described elsewhere\cite{khaliullin2007,horn2016probing,Mao:2021:EDA-review}, but we provide two important clarifications for this study. First the most appropriate choice for $E_{\rm elec}$ is the quasi-classical expression, which depends only on the geometry of individual monomers\cite{mao2017energy}, and we use the fragment electric response function approach (at the dipole plus quadrupole level) to evaluate the polarization, ensuring a well-defined basis set limit.\cite{horn2015} Note that we will use a convention of referring to all energy terms in the force field with a $V$ and all energy terms from electronic structure with an $E$. Finally, a detailed description of the water data generation and parameterization procedure is given in Supplementary Information.

\subsection*{Permanent Electrostatics and Dispersion}
\textbf{Electrostatics and charge penetration}. Our description of electrostatics comes from a traditional point multipole approach up to the quadrupoles, and a charge penetration (CP) contribution that modifies the short-range electrostatic energy to be more attractive than the point multipole expansion alone. We isolate the CP energy by taking the total classical electrostatic energy from EDA minus the point multipole interaction energy when using Stone's distributed multipole analysis (DMA)\cite{stone1981distributed,stone1985distributeda} out to hexadecapoles on all atoms.

\begin{equation}
  E^{CP}=E^{elec}_{EDA}-E^{elec}_{DMA}
  \label{eq:cp}
\end{equation}
The advantage of this approach is it allows us to ensure that our multipoles are not biased to compensate for error in the description of charge penetration, and vice versa, which is essential to reproduce the classical electrostatic energy in EDA.

CP is described by treating each atom as having both a positively charged core and negatively charged shell. Considering the interactions of the collection of cores and shells, which are expanded in multipoles, results in the following electrostatic energy expression:
\begin{equation}
  V_{elec}=\sum_{i<j}Z_iT_{ij}Z_j+Z_i\bm{T}_{ij}^{damp}\bm{M}_j+Z_j\bm{T}_{ji}^{damp}\bm{M}_i+\bm{M}_i\bm{T}_{ij}^{overlap}\bm{M}_j
  \label{eq:elec}
\end{equation}
The first term in Eq. \ref{eq:elec} represents repulsive core-core interactions where $T_{ij}=1/r_{ij}$ with $Z_i$ the core charge on the $i$th atom; note that this is not the nuclear charge but an effective nuclear charge. The second and third terms describe attractive core-shell interactions where $\bm{M}_i$ is a vector whose entries are the components of the multipoles located on that atom. The final term corresponds to the shell-shell interactions. The core-shell and shell-shell interaction tensors are defined in Eqs. \ref{eq:tensor_1} and \ref{eq:tensor_2}.\\

\noindent
\textbf{Dispersion}. The dispersion energy uses a damped polynomial interaction given by,
\begin{equation}
  V_{disp}=\sum_{i<j}f_6^{TT}(r_{ij})\frac{C_{6,ij}}{r_{ij}^6}
  \label{eq:disp}
\end{equation}
\noindent
where $C_{6,ij}$ is the dispersion coefficient between atoms $i$ and $j$ which is determined as $C_{6,ij}=\sqrt{C_{6,i}C_{6,j}}$, and $C_{6,i}$ is a parameter fit to the EDA dispersion energy. $f_6^{TT}(r_{ij})$ is the sixth-order Tang-Toennies (TT) damping function\cite{tang1984improved} which was originally derived to damp short-range dispersion,
\begin{equation}
  f_n^{TT}(r_{ij}) = 1-e^{-r_{ij}}\sum_{k=0}^n\frac{r_{ij}^k}{k!}
  \label{eq:TT}
\end{equation}
\noindent
In their original work, Tang and Toennies show that the appropriate choice of $n$ for dispersion is $n=6$. This makes the damping function an exponential multiplied by a sixth order polynomial. This polynomial is able to control the $r^{-6}$ scaling of dispersion, while the exponential ensures no damping at long distances.

Interestingly, the $n=6$ Tang-Toennies damping function is exactly the same as the dipole-quadrupole damping function generated from $f_{overlap}$ in Eq. \ref{eq:f_overlap}. The reason the appropriate damping function derived from the density occurs at the dipole-quadrupole rather than dipole-dipole level is that at short-range all attractive electrostatic interactions have a $-1/r$ asymptote. This divergence is cancelled out by the $1/r$ nuclear-nuclear repulsion. For dispersion, however, the damping function itself needs to eliminate the short-range divergence. This was considered by Tang and Toennies by enforcing that the dispersion energy becomes zero at the united atom limit. The fact that a Slater density generates the Tang-Toennies damping function also indicates Eq. \ref{eq:slater} is a good choice of model density.

\subsection*{Pauli Repulsion, Polarization, Charge Transfer}
Here we highlight the unique aspects of the CMM model through the introduction of new physics and functional forms to describe Pauli repulsion, exchange and many-body polarization, and many-body charge transfer flow. Polarization is handled in a manner that allows for both intramolecular charge fluctuations and induced dipoles. At short range, exchange polarization becomes non-negligible and is handled by a short-range multipole expansion. Charge transfer is also described by a short-range multipole expansion, allowing for intramolecular charge fluctuations and explicit charge transfer between molecules, and is formulated to naturally describe many-body charge transfer. Pauli repulsion is modeled exclusively by a repulsive short-range multipole expansion.

All multipole expansions used by Pauli, polarization, and CT share the same anisotropy of the electric multipoles described in the previous section. This then requires only a single parameter at each rank of the expansion, which we refer to as $K_q$, $K_\mu$, and $K_\Theta$ as the CMM model truncates the expansion at quadrupoles, although it is straightforward to include higher order multipoles if desired. While this functional form constraint is an approximation, as we show below the anisotropy of electrical moments is an excellent shape basis for all EDA terms. More importantly, it dramatically decreases the number of parameters, circumventing the need to fit many independent multipole expansions by EDA term, while  gaining computational efficiency as well.  \\

\noindent
\textbf{Pauli Repulsion.} The importance of anisotropy in Pauli repulsion and its connection to classical electrostatics have recently been highlight by Rackers \textit{et al.}.\cite{rackers2019classical,rackers2021polarizable,chung2024beyond} We note that our approach of representing  Pauli repulsion in terms of multipoles has an interesting physical interpretation. Namely, as two electron densities begin to overlap, the electrons will be expelled from the internuclear region in order to keep the total system wavefunction antisymmetric. This results in a "hole" in the electron density where nuclei are exposed to one another.\cite{rackers2019classical} Hence the multipoles describe the magnitude and shape of the depletion of electron density between two atoms which are near one another. 

But because the electric multipoles arise from an expansion in $1/r_{ij}$, while Pauli repulsion has no long-range contribution and scales at least exponentially at short-range, the functional forms based on classical electrostatics need modification. The CMM resolves this issue by using the difference of an undamped and damped interaction tensor $\mathbf{T}_{ij}^{sr,+}$ as defined in Eq. \ref{eq:mult_sr_plus}. This allows for anisotropy to be included in Pauli repulsion while rigorously eliminating the long-range $1/r_{ij}$ interaction. Thus we define the Pauli repulsion energy as,
\begin{equation}
    V_{Pauli}=\sum_{i<j}\mathbf{M}_i^{Pauli}\mathbf{T}_{ij}^{sr,+}\mathbf{M}_j^{Pauli}
    \label{eq:pauli}
\end{equation}
\noindent
where $\mathbf{M}_i^{Pauli}$=$\mathbf{K}_i\mathbf{M}_i^{Elec}$, is a vector representation of all multipoles up to quadrupoles. As stated above, we explicitly fit only the Pauli charges, $K_{q}$, while forcing the Pauli dipoles, $K_{\mathbf{\mu}}$, and quadrupoles, $K_{\mathbf{\Theta}}$, to be proportional to the electric dipoles and quadrupoles. 

When we introduce analysis of data generated from force decomposition analysis (FDA)\cite{aldossary2023force}, we find that there is a force along \ce{O-H} bonds not directly captured by Eq. \ref{eq:slater}. We thus make the Pauli repulsion charges, $K_i^q$, dependent on the \ce{O-H} bond length:
\begin{equation}
K_i^q(R_{ij})=K_i^q+j_{b,pauli}^{\mathrm{OH}}(R_{ij}-R_e)
  \label{eq:variable_pauli}
\end{equation}
to recover this non-neligible effect on the intramolecular degrees of freedom.\\

\noindent
\textbf{Polarization}. While distributed polarization naturally contains both charge-flow and induced dipole contributions\cite{stone1985distributedb}, typically the charge-flow contributions are eliminated through localization.\cite{ruth1994localization} Our approach allows for charge flow polarization using a modification of the electronegativity equalization model (EEM).\cite{mortier1986electronegativity} In EEM, the energy of a molecule is expanded to second-order as a function of charge while allowing all charges to interact
\begin{equation}
  V(\bm{q})=\sum_i \chi_i q_i + \frac12 \sum_i \eta_i q_i^2 + \sum_{i<j} \frac{q_i q_j}{r_{ij}}
  \label{eq:eem}
\end{equation} where $\chi_i$ represents the electronegativity of atom $i$ and $\eta_i$ is the atomic hardness of atom $i$. By requiring the electronegativity of all atoms to become become equal, new atomic charges are determined by solving a system of linear equations. 

There are several known shortcomings of EEM for non-reactive FFs including unphysical long-range transfer of charge between molecules\cite{chen2007qtpie,chen2008unified,Leven2021} as well as a change in charge of atoms in a molecule that interferes with the definition of the permanent electrostatics. Our solution to the first problem is to allow charge rearrangements within a molecule but not between molecules which is enforced by the method of Lagrange multipliers. Although this constraint is relaxed below for charge transfer, it is designed strictly to satisfy agreement with the polarization term of EDA. For the second problem, we drop the linear term in Eq. \ref{eq:eem} and focus only on the fluctuation of charge around the reference charge used for the permanent electrostatics. Thus we are equalizing electronegativity around an "already equalized" state, and the change in electronegativity at each atom due to an environment is simply the electric potential at that atom. We can then write the fluctuating charge (FQ) contribution to the energy as,
\begin{equation}
  V(\delta \bm{q})=\frac12\sum_i \eta_i \delta q_i^2 + \sum_i \delta q_i V_i + \sum_{i<j}\frac{\delta q_i \delta q_j}{r_{ij}} + \sum_{\alpha}\lambda_\alpha \sum_{i\in\alpha}\delta q_{i}
  \label{eq:fq}
\end{equation}

We also allow electric fields due to the environment to induce dipoles on all atoms as done previously for other polarization models.\cite{Demerdash2014,Demerdash2018} The energy of an induced dipole $\bm{\mu}_i^{ind}$ in an electric field, $\bm{E}$, including mutual polarization is,
\begin{equation}
  V(\bm{\mu}^{ind})=-\frac12\sum_i \bm{\mu}_i^{ind}\cdot \bm{E}_i^{overlap} + \sum_{i<j}\bm{\mu}^{ind}_i \bm{T}^{pol}_{ij,\mu\mu}\bm{\mu}^{ind}_j
  \label{eq:induced_dipoles}
\end{equation}
The field $\bm{E}_i^{overlap}$ is the damped electric field generated by two overlapping Slater densities at atom $i$. $\bm{T}^{pol}_{ij,\mu\mu}$ is the dipole-dipole interaction tensor which is derived from appropriate gradients of $f_{ij}^{pol}/r_{ij}$, where $f_{ij}^{pol}$ is a damping function specifically for mutual polarization. However, the damping functions generated from Eq. \ref{eq:f_overlap} have a polynomial that is one order too small to control the short-range singularity for mutual polarization. We therefore extend the polynomial by one order according to
\begin{equation}
    f_1^{pol}(r_{ij})=1 - \left(1 + \frac{1}{9}(b_{ij}r_{ij}) + \frac{1}{11}(b_{ij}r_{ij})^2 + \frac{1}{13} (b_{ij}r_{ij})^3 + \frac{1}{15}(b_{ij}r_{ij})^4\right)e^{-b_{ij}r_{ij}}
    \label{eq:f1_pol}
\end{equation}
\noindent
and the $f_3^{pol}$ and $f_5^{pol}$ are generated from gradients of $f_1^{pol}/r_{ij}$. This results in a sixth-order polynomial in the dipole-dipole interaction tensor which is sufficient to eliminate the short-range dipole-dipole singularity. Hence the form of the $ij$ entries of the mutual polarization interaction tensors are 
\begin{subequations}
  \begin{equation}
    T^{pol}_{ij,qq}=f_1^{pol}\frac{1}{r_{ij}}
    \label{eq:tensors_a}
  \end{equation}
  \begin{equation}
  \bm{T}^{pol}_{ij,q\mu}=f_3^{pol}\frac{-\bm{r}_{ij}}{r_{ij}^3}
    \label{eq:tensors_b}
  \end{equation}
  \begin{equation}
\bm{T}^{pol}_{ij,\mu\mu}=\left(f_5^{pol}\frac{\bm{r}_{ij}\otimes\bm{r}_{ij}}{r_{ij}^5}-f_3^{pol}\frac{\bm{1}}{r_{ij}^3}\right)
    \label{eq:tensors_c}
  \end{equation}
  \label{eq:tensors}
\end{subequations}

What now remains is to determine the values of $\delta \bm{q}$ and $\bm{\mu}^{ind}$ which minimize the total energy of the system. To do this we take the derivative with respect to each $\delta q_i$ and each component of each $\bm{\mu}_i^{ind}$ and set them all equal to zero. This results in a system of linear equations which can be written succinctly as follows:
\begin{equation}
  \begin{pmatrix}
    \bm{T}^{qq} & \bm{1}_\lambda & \bm{T}^{q\mu} \\
    \bm{1}_\lambda^\dagger & 0 & 0 \\
    -\bm{T}^{\mu q} & 0 & \bm{T}^{\mu\mu} \\
  \end{pmatrix}
  \begin{pmatrix}
    \delta \bm{q} \\
    \bm{\lambda} \\
    \bm{\mu} \\
  \end{pmatrix}
  =
  \begin{pmatrix}
    -\bm{V} \\
    \bm{Q} \\
    \bm{E} \\
  \end{pmatrix}
  \label{eq:pol_mat}
\end{equation}
where $\delta\bm{q}$ contains the optimally rearranged charges, $\bm{\lambda}$ are the Lagrange multipliers which enforce charge conservation, and $\bm{\mu}$ are the induced dipoles. The solution vector in Eq. \ref{eq:pol_mat} contains the electric potential, $\bm{V}$, the total charges of each molecule, $\bm{Q}$, and the electric field on each atom $\bm{E}$. The matrix has several blocks containing the charge-charge ($\bm{T}^{qq}$), charge-dipole ($\bm{T}^{q\mu}$), dipole-charge ($\bm{T}^{\mu q}$), and dipole-dipole interaction tensors ($\bm{T}^{\mu\mu}$). Note that the diagonal elements of $\bm{T}^{qq}$ are the atomic hardness $\eta$ and the $3\times 3$ diagonal blocks of $\bm{T}^{\mu\mu}$ are the inverse polarizability tensor $\bm{\alpha}_i^{-1}$. The block $\bm{1}_\lambda$ has a column for each molecule in the system, with an entry of 1 if the $i$th atom is in that molecule and 0 otherwise, that enforces the charge-conservation constraints for each molecule. An extension to include quadrupole polarization can be achieved by adding additional blocks in an analogous manner to charges and dipoles, in which quadrupoles would be induced by the field gradient at each atom.

Finally, we define a term which describes the coupling between exchange and polarization, and which scales exponentially with distance.\cite{patkowski2020recent,Mao:2021:EDA-review} The Exch-Pol energy, $V_{exch-pol}$, is 
\begin{equation}
V_{exch,pol}=\sum_{i<j}\mathbf{M}_i^{exch,pol}\mathbf{T}_{ij}^{sr,-}\mathbf{M}_j^{exch,pol}
    \label{eq:exch_pol}
\end{equation}
\noindent
where $\mathbf{M}_i^{exch,pol}$ are the exchange-polarization multipoles which interact with one another via the attractive short-range multipole interaction tensor $\mathbf{T}_{ij}^{sr,-}$ defined in Eq. \ref{eq:mult_sr_minus}. We have written Eq. \ref{eq:exch_pol} in its multipolar form to emphasize its generality, but we only include the rank zero term for water.\\

\noindent
\textbf{Charge Transfer.} The true definition of charge transfer involves the movement of charge density between molecules\cite{thirman2018characterizing}, and involves both attractive and repulsive contributions. The attractive component is the energy lowering associated with delocalizing the electron density, while the repulsive contribution arises from any molecule having a non-integer total charge. Quantum mechanically, the energy penalty is associated with partial occupation of an anti-bonding orbital.
Charge transfer has historically been the most difficult of the terms in EDA to model since there is no classical analogue to the QM charge transfer process involving electron flow.\cite{herman2023accurate}  Hence we introduce a new approach to describing direct CT and many-body CT which is enabled by the fact we allow for explicit charge rearrangements in our description of polarization. 

The direct contributions allow for energetic stabilization associated with both forward and backward CT, and are described by an attractive short-range multipole expansion $\mathbf{T}_{ij}^{sr,-}$ as defined in Eq. \ref{eq:mult_sr_minus}.
\begin{subequations}
  \begin{equation}
V_{CT}^{i\rightarrow j}=\sum_{i<j}\mathbf{M}^{i\rightarrow j}_{CT} \mathbf{T}_{ij}^{sr,-}\mathbf{M}^{i\rightarrow j}_{CT}
\end{equation}
\begin{equation}
  V_{CT}^{j\rightarrow i}=\sum_{i<j}\mathbf{M}^{j\rightarrow i}_{CT} \mathbf{T}_{ij}^{sr,-}\mathbf{M}^{j\rightarrow i}_{CT}
\end{equation}
\begin{equation}
  V_{CT}^{direct}=\sum_{i<j}V_{CT}^{i\rightarrow j}+V_{CT}^{j\rightarrow i}
\end{equation}
  \label{eq:ct_direct}
\end{subequations}
\noindent
The short-range multipole interactions for CT are unique because, in accordance with EDA, both forward and backward charge transfer are possible and hence there are two sets of multipoles associated with CT. For the CMM water model we choose to go up to rank two for the donor multipoles and only rank zero for the acceptor multipoles. The motivation for this is that charge is most often transferred out of orbitals containing lone pairs which are usually quite anisotropic.

We take inspiration from perturbation theory which shows that, to a good approximation, the amount of charge transferred between two molecules is proportional to the energy associated with forward and backward CT.\cite{khaliullin2007,khaliullin2008analysis,khaliullin2009electron} Therefore, we define the amount of charge transferred from $i$ to $j$, $\Delta Q^{CT}_{i\rightarrow j}$, and from $j$ to $i$, $\Delta Q^{CT}_{j\rightarrow i}$, as
\begin{subequations}
  \begin{equation}
  \Delta Q_{CT}^{i\rightarrow j}=\frac{V_{CT,iso}^{i\rightarrow j}}{\epsilon_{i\rightarrow j}}
\label{eq:ct_forward}
\end{equation}
\begin{equation}
  \Delta Q_{CT}^{j\rightarrow i}=\frac{V_{CT,iso}^{j\rightarrow i}}{\epsilon_{j\rightarrow i}}
\end{equation}
\end{subequations}
The proportionality constant between the direct CT energy and the amount of transferred charge is written as $\epsilon_{i\rightarrow j}$ to emphasize that this proportionality is related to the difference in energy of an occupied orbital on $i$ and an unoccupied orbital on $j$.\cite{khaliullin2007} We choose this to be a pair-specific parameter since it avoids having to choose an arbitrary combination rule, and it decreases the total number of parameters because many pairs do not exchange appreciable amounts of charge. For instance, in water, only the oxygen to hydrogen parameter is relevant. Note also that $V_{CT,iso}^{i\rightarrow j}$ is the isotropic contribution to the charge transfer potential in Eq. \ref{eq:ct_direct}. We use just the isotropic part since the anisotropic part makes little contribution to the amount of charge transferred, and it also simplifies the derivatives.

Because we allow charge to explicitly move between fragments for CT, we  modify the molecular charge constraints used in Eq. \ref{eq:pol_mat}. The charge constraint for fragment $A$ will now take the form,
\begin{equation}
  Q_{CT}^A=Q_A+\sum_{i\in A}\sum_{j\notin A}\Delta Q_{CT}^{j\rightarrow i}-\Delta Q_{CT}^{i\rightarrow j}
  \label{eq:charge_constraint}
\end{equation}
in which $Q_{CT}^A$ is the difference in charge transferred to atom $i$ (in $A$) and charge transferred from atom $i$, summed over all atoms in molecule $A$. These charges will not be optimally distributed, so they will be allowed to relax during the polarization process. This allows us to capture the so-called "re-polarization"\cite{khaliullin2007} effect in which orbitals relax after allowing for occupied-virtual mixing. For example, when charge is transferred from oxygen to hydrogen in a water dimer, the final excess charge will mostly come to rest on the oxygen in the water with net-negative charge. This re-polarization gives rise to the indirect contributions to charge transfer,
\begin{equation}
  V^{indirect}_{CT}=V_{pol}(Q_{CT})-V_{pol}(0)
  \label{eq:ct_indirect}
\end{equation}
\noindent
The non-additive contribution to CT is defined as the polarization energy with CT, $V_{pol}(Q_{CT})$, minus the polarization energy without CT, $V_{pol}(0)$. Because the charge transferred between fragments is proportional to the direct CT contributions, the charge constraints depend on the distance between atoms. This means there is a gradient contribution which multiplies the Lagrange multipliers with the gradient of $\Delta Q_{CT}^{i\rightarrow j}$ and $\Delta Q_{CT}^{j\rightarrow i}$. This is not difficult or expensive to evaluate, but it is an unusual gradient term which must be accounted for in software implementations.

\subsection*{One-Body Potential} 
The deformation energy for a single water molecule is constructed following a protocol we have recently published for any molecule without soft torsions.\cite{Sami2024} The one-body potential consists of a Morse potential, cosine angle potential, a bond-bond coupling term, and bond-angle coupling term.
\begin{align}
  \label{eq:morse}
  V_{bond}&=D_{\mathrm{OH}}\left[ 1-\exp(-\alpha (R-R_e))\right]^2 \\
  \label{eq:bond_bond}
  V_{bb}&=k_{bb}(R_1-R_e)(R_2-R_e) \\
  \label{eq:angle}
  V_{angle}&=\frac{k_a}{2}(\cos\theta-\cos\theta_e)^2 \\
  \label{eq:bond_angle}
  V_{ba}&=k_{ba}(R-R_e)(\cos\theta-\cos\theta_e)
\end{align}
where $D_{\mathrm{OH}}$ is the dissociation energy of the \ce{O-H} bond in water, $R_e$ is the equilibrium bond length in water, and $\alpha=\sqrt{k_e/2D}$ determines the curvature of the potential as is evident from the fact it is written in terms of the harmonic force constant, $k_e$. The two \ce{O-H} stretches in water are coupled linearly in Eq. \ref{eq:bond_bond} via a single bond-coupling parameter, $k_{bb}$. The angle potential is harmonic in $\cos\theta$ where $\theta$ is the \ce{HOH} angle and $\theta_e$ is the equilibrium angle in water, as seen in Eq. \ref{eq:angle}, and  Eq. \ref{eq:bond_angle} shows that the angle and bond potentials are linearly coupled by a single parameter, $k_{ba}$. The parameters are fit to reproduce the CCSD(T)/aug-cc-pV5Z Hessian at the corresponding equilibrium geometry using the Q-Force package\cite{Sami2021,Witek2023}. Note that this is the only term for which we do not use $\omega$B97X-V/def2-QZVPPD as a reference, simply because CCSD(T)/aug-cc-pV5Z is closer to the experimental water monomer geometry.

As will be shown, the polarizability derivatives of the water monomer are not possible to reproduce using just atomic dipole polarizabilities. Our model, however, includes fluctuating charges which improve the polarizability derivatives considerably, although the agreement with polarizability derivatives computed from electronic structure is still flawed. In the same way the dipole derivatives can be reproduced accurately by including charge flux in a model\cite{Dinur1995, liu2019implementation}, we have implemented geometry-dependent atomic hardness parameters, $\eta$.
\begin{equation}
  \eta_{\mathrm{H1}} = \eta_\mathrm{H} \left(\frac{R_e}{R_{\mathrm{OH,1}}}\right)^{k^\eta}\left(\frac{R_e}{R_{\mathrm{OH,2}}}\right)^{k^\eta_{bb}} + k^\eta_{a}(\theta - \theta_e)
  \label{eq:variable_hardness}
\end{equation}
In Eq. \ref{eq:variable_hardness}, the atomic hardness of a particular hydrogen, $\eta_{\mathrm{H1}}$, is modified based on the length of both \ce{O-H} bonds, $R_{\mathrm{OH,1}}$ and $R_{\mathrm{OH,2}}$, and the angle $\theta$. The parameters $k^\eta$, $k^\eta_{bb}$, and $k^\eta_{a}$ describe the magnitude of change in atomic hardness and are fit to reproduce the polarizability derivatives computed from electronic structure. This particular functional form was chosen to be well-behaved when either bond is elongated, such that when the hardness of atom $\mathrm{H1}$ is decreased, it increases the polarizability along that bond. This is a source of so-called electrical anharmonicity and hence contributes to the large, positive second dipole derivative associated with hydrogen-bonded water molecules.\cite{mccoy2014role} Note that this term adds negligible cost to the force field evaluation since we already compute the derivatives with respect to each internal coordinate when computing the deformation energy.

Like electrostatics, polarization parameters need to be constrained to give physically meaningful parameters. Specifically, in addition to EDA energies, we include the polarizability and polarizability derivatives at the $\omega$B97X-V/def2-QZVPPD equilibrium geometry of water in the fitting process. The loss function we minimize against is,
\begin{equation}
  L_{pol}=\sqrt{\frac{\sum_{i=1}^{N}(V_i^{FF}-E_i^{EDA})^2}{N}} + w_1||\bm{\alpha}^{FF}-\bm{\alpha}^{EDA}||+w_2||\frac{\partial\bm{\alpha}^{FF}}{\partial \bm{r}}-\frac{\partial\bm{\alpha}^{EDA}}{\partial \bm{r}}||
\label{eq:pol_loss}
\end{equation}
\noindent
In the above, the first term is the RMSD of the predicted energies, $V_i^{FF}$, from the EDA energies $E_i^{EDA}$. The second term is the Frobenius norm of the difference between the computed and predicted molecular polarizabilities, $\bm{\alpha}$. The third term is the same as the second but for the polarizability derivatives. The weights, $w_1$ and $w_2$ are set to 1.0 and 0.5 respectively. This, in essence, forces the molecular polarizability to be reproduced exactly while allowing for some error in the polarizability derivatives which are much more difficult to reproduce.

In the CMM model the intramolecular polarization is described by coupling the bonding potential to the environment through the electric field, which makes the polarization energies more accurate and dramatically improves the underlying forces. Furthermore, this term enables us to more accurately reproduce the well-known structure-frequency correspondence in water.\cite{boyer2019beyond} We do this by modifying Eq. \ref{eq:morse} to also be dependent on the environment\cite{boyer2019beyond} by coupling $k_e$ and  $R_e$ to the electric field projected along the bond, $\mathrm{E_{OH}}$, via the first and second dipole derivatives, $\mu^{(1)}$ and $\mu^{(2)}$. The equilibrium bond length, $R_e$, becomes
\begin{equation}
  R_e(\mathrm{E_{OH}})=R_e^0+ \frac{\mathrm{E_{OH}}\mu^{(1)}}{k_e^0-\mathrm{E_{OH}}\mu^{(2)}}
  \label{eq:bond_in_field}
\end{equation}
where $R_e^0$ is the equilibrium bond length and $k_e^0$ is the force constant under zero field. The force constant under a nonzero field, $k_e(\mathrm{E_{OH}})$, is 
\begin{equation}
  k_e(\mathrm{E_{OH}})=k_e^0-3k_e^0\alpha\left(R_e(\mathrm{E_{OH}})-R_e^0\right)-\mathrm{E_{OH}}\mu^{(2)}
  \label{eq:force_constant_in_field}
\end{equation}
These equations can be derived by analyzing the behavior of a Morse potential in an electric field\cite{boyer2019beyond} and guarantees the structure-frequency correlation will be respected at least approximately. The dipole derivatives needed to evaluate the field-dependent Morse potential, Eqs. \ref{eq:bond_in_field} and \ref{eq:force_constant_in_field}, are computed from electronic structure by scanning along the \ce{O-H} bond length of a water monomer, although the model accurately predicts the parameters from its own dipole surface. The molecular dipole moment, $\bm{\mu}$, is projected along the \ce{O-H} bond unit vector, $\hat{R}_{\mathrm{OH}}$, to give $\mu_{\mathrm{OH}}=\bm{\mu}\cdot\hat{R}_{\mathrm{OH}}$. $\mu_{\mathrm{OH}}$ is then fit to a second-order polynomial whose coefficients directly give the dipole derivatives (see Supplementary Figure S1).

By using the above equations we will obtain the correct slope of the structure-frequency correlation, but it will tend to underestimate the actual bond length and frequency shifts. The final step to reproduce the bond lengths and vibrational frequencies is to include a contribution from CT. This is motivated by adiabatic EDA calculations, where the largest contribution to bond elongation and red-shifting occurs on the CT surface.\cite{mao2017energy} Specifically, we allow both the bond length and force constants to be modified according to the amount of charge transferred into a hydrogen atom, as computed with Eq. \ref{eq:ct_forward}. This results in the final expressions used for the bond length and force constant in our environment-dependent bonding potential,
\begin{align}
  \label{eq:bond_in_field_2}
  R_e(\mathrm{E_{OH}})&=R_e^0+ \frac{\mathrm{E_{OH}}\mu^{(1)}}{k_e^0-\mathrm{E_{OH}}\mu^{(2)}} + k_{ct}^{(1)}(\Delta Q_{CT}^{\mathrm{H}})^2 \\
  \label{eq:force_constant_in_field_2}
  k_e(\mathrm{E_{OH}})&=k_e^0-3k_e^0\alpha\left(R_e(\mathrm{E_{OH}})-R_e^0\right)-\mathrm{E_{OH}}\mu^{(2)}+k_{ct}^{(2)}(\Delta Q_{CT}^{\mathrm{H}})^2
\end{align}
The rationale for using the amount of charge accepted by a hydrogen atom, $\Delta Q_{CT}^{\mathrm{H}}$, in Eqs. \ref{eq:bond_in_field_2} and \ref{eq:force_constant_in_field_2} is that this charge is transferred into an anti-bonding orbital and hence should have a large effect on the \ce{O-H} bond in question. Similarly, we ignore the charge transferred out of the oxygen atom since these electrons come from non-bonding orbitals and should therefore minimally affect the \ce{O-H} bonds in that water. This introduces two additional parameters, $k_{ct}^{(1)}$ and $k_{ct}^{(2)}$, which determine the sensitivity of the bond-length and force constant to CT. The importance of these parameters is that while they have a very small effect on the CT energy, they contribute a sizable effect on the CT forces. In the Supplementary Information, we illustrate how the environment-dependent bonding potential improves energies and forces for the water dimer.

\section*{Results}
\subsection*{Water Monomer Properties}
In the construction of the CMM model, we wish to ensure that the model reproduces as many properties of the water monomer as possible. Here we compare the dipole surface, molecular polarizability, and polarizability derivatives of the CMM model against the $\omega$B97X-V/def2-QZVPPD reference. By ablating certain interactions, we can trace the origins for CMM improvement for each of the monomer properties. It was first pointed out by Fanourgakis and Xantheas that reproducing the dipole surface of water is essential for capturing the opening of the bend angle from $104.5^{\circ}$ for water clusters as they become larger and ultimately to $~106.5^{\circ}$ for the condensed phase.\cite{fanourgakis2006flexible} 

\begin{figure}[H]
  \includegraphics*[width=0.99\textwidth]{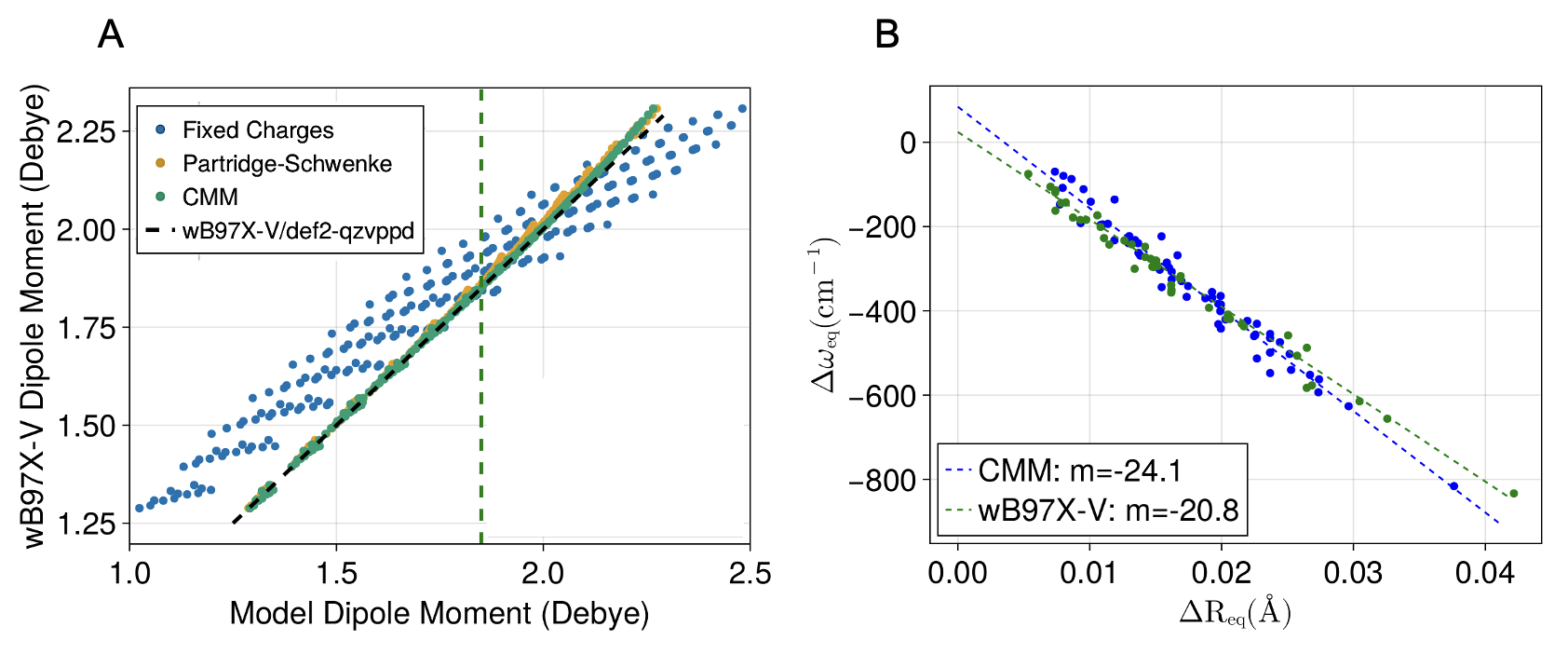}
  \caption{\textit{Comparison of Models for Water Monomer Properties}. (A) The dipole surface of water of various models for all structures with a deformation energy less than 20 kcal/mol. The black dashed line shows the values computed with $\omega$B97X-V/def2-QZVPPD. The green dashed line corresponds to the experimental gas-phase dipole moment
  of water of 1.85 Debye. (B) Correlation of $\Delta\omega$ vs $\Delta R_e$ over a collection
  of low-energy structures of \ce{(H_2O)_{2-6}} using CMM and $\omega$B97X-V/def2-QZVPPD. The linear fits are not constrained to pass through zero
  which explains the slightly large slopes compared to previous work.
}
  \label{fig:onebody}
\end{figure}

In Figure \ref{fig:onebody}A, we make a comparison between the dipole surface of the CMM model (green), a dipole surface with fixed charges and dipoles that optimally reproduce the EDA electrostatic energy of the water dimer (blue), the dipole surface using the Partridge-Schwenke (PS) model\cite{partridge1997determination} (orange), along with the reference dipole surface computed with $\omega$B97X-V/def2-QZVPPD. Clearly, fixed charge force fields completely fail to reproduce the dipole surface of water whereas the dipole surface associated with the PS water monomer surface is exact over a wide range of energies by construction.\cite{partridge1997determination} Figure \ref{fig:onebody}A shows that the CMM dipole surface is as good as that for PS, reproducing the dipole derivatives of water at its equilibrium geometry to five decimal places in atomic units.

We evaluate the polarizability of the CMM for a water molecule with the y-axis as the bisector of the HOH angle and the z-axis normal to the plane of the water molecule;  the model reproduces the $\omega$B97X-V/def2-QZVPPD molecular polarizability up to three decimal places in bohr$^3$ as seen in Table \ref{tab:pol_derivs}. The polarizability derivatives control the intensity of peaks measured with Raman spectroscopy, and additionally, indicates how well the molecular polarizability at distorted geometries will be reproduced, but this property is rarely considered in the construction of water models.\cite{sidler2018efficient} Table \ref{tab:pol_derivs} reports the polarizability derivatives of the CMM model as well as the polarizability derivatives for an identically parameterized model which does not include fluctuating charges, and compare them to the QM reference model. Because charge fluctuations for water contribute to in-plane polarization, we find that the CMM gives much better xx, xy, and yy polarizability derivatives than one which just uses anisotropic dipole polarizabilities. Note that the polarizability derivatives in Table \ref{tab:pol_derivs} are also improved considerably by the geometry-dependent atomic hardness described in Eq. \ref{eq:variable_hardness}. This indicates that one of the main reasons water models have historically predicted Raman intensities very poorly\cite{hamm20142d} is the lack of fluctuating charges in the polarization process. 

\begin{table}[ht!]
  \begin{center}
  \begin{tabular}{ccccc}
      \multicolumn{5}{c}{Molecular Polarizability and Polarizability Derivatives of Water} \\\hline
      Molecular (bohr$^3$) & $\alpha_{xx}$ & $\alpha_{yy}$ & $\alpha_{zz}$ & \\\hline
                  & 10.0321 & 9.65958 & 9.40921 & \\\hline
       Atom (bohr$^2$) & $xx$ & $xy$ & $yy$ & $zz$ \\\hline
       O$x$ & -                 & 4.04/4.05/-0.13    & -                  & -  \\
       O$y$ & 5.15/5.26/2.04    & -                  & 4.45/4.22/-2.03    & 1.50/0.0/0.0  \\
       O$z$ & -                 & -                  & -                  & -  \\
       H$x$ & -4.61/-4.79/0.78  & -2.02/-2.02/0.06   & -2.53/-2.38/-0.78  & -1.39/0.0/0.0  \\
       H$y$ & -2.57/-2.63/-1.02 & -1.68/-1.68/-0.08  & -2.22/-2.11/1.02   & -0.75/0.0/0.0  \\
       H$z$ & -                 & -                  & -                  & -  \\\hline
  \end{tabular}
  \end{center}
  \vspace{-3mm}
  \caption{\textit{Molecular polarizability and polarizability derivatives of water computed with and without  fluctuating charges as compared to DFT.} The water monomer has its bisector aligned with the $y$-axis and the $z$-axis is normal to the water molecule plane. For polarizability derivatives, the first entry is computed with $\omega$B97X-V/def2-QZVPPD, the second with CMM, and the third with the same model but using parameters optimized without fluctuating charges. Note that the derivatives for the second hydrogen are identical to the first but with opposite sign. The $xz$ and $yz$ entries are omitted since they are small and reproduced to three decimal places by both FF models.}
  \label{tab:pol_derivs}
\end{table}

While the in-plane polarizability derivatives of CMM are reproduced very accurately, the $zz$ polarizability derivatives are not as well described. The $zz$ polarizability derivatives of water are an interesting case since they can be reproduced accurately when using intramolecular induced dipole interactions, but this tends to make the polarization energies worse and will make polarization catastrophes more likely for ion-water systems, for example. We can also reproduce the $zz$ polarizability derivatives by allowing the z-component of atomic polarizabilities to be geometry-dependent. We decided not to do this since it adds additional complexity with a fairly small benefit for water, but for a system like benzene which has strong $\pi$-$\pi$ interactions we will take that additional step to recover accurate out-of-plane polarizability derivatives in the future.

In order for a FF to be useful for theoretical spectroscopy, it must respect the relationships between structure and vibrational frequencies. In the case of water, this manifests as a linear relationship between the change in bond length ($\Delta R_e$) and change in \ce{O-H} stretching frequency ($\Delta\omega$)\cite{boyer2019beyond}, also known as Badger's rule, with a slope of $\approx -20\ \mathrm{cm^{-1}}/$0.001\AA \hspace{1pt} when evaluated across a large data set.\cite{badger1934relation} 
Figure \ref{fig:onebody}B shows that while the DFT reference model yields a reasonably accurate slope, Boyer \textit{et al.} found that a field-independent Morse oscillator with parameters appropriate to water yields a reduced slope of $\approx -11\ \mathrm{cm^{-1}}/$0.001 \hspace{1pt}\AA.  Thus to obtain a correct slope requires an  electric field dependence\cite{boyer2019beyond}, and motivates our bonding potential parameters to be modulated by the field along the O-H bond, as given by  Eqs. \ref{eq:bond_in_field} and \ref{eq:force_constant_in_field}. We computed the necessary dipole derivatives from a simple \ce{O-H} scan and found the parameters $\mu^{(1)}=0.1654$ and $\mu^{(2)}=-0.01246$. If we do the same calculation with CMM, we get $\mu^{(1)}=0.1658$ and $\mu^{(2)}=-0.0204$. This indicates that as long as a model has an accurate dipole surface, the dipole derivatives needed to compute the field-dependence of a Morse potential can be computed directly from the force field. The final result of the structure-frequency correlation in water using the CMM model is shown in Figure \ref{fig:onebody}B. We consider this an excellent result given the simplicity of the field-dependent Morse potential, especially since it requires no free parameters.

\subsection*{Water Intermolecular Interactions}
The CMM model has a very accurate dimer surface, even in relatively high-energy configutations, as demonstrated in Figures \ref{fig:w2_scans}A and \ref{fig:w2_scans}B. Figure \ref{fig:w2_scans}A shows each component of the EDA energy as computed with CMM and $\omega$B97X-V for a scan over the \ce{O-O} distance of the water dimer. CMM reproduces the long-range asymptotics of each term, but more importantly the agreement is excellent in the short-range, only showing deviations deep in the overlapping region. Perhaps even more impressive is the agreement of each term over the full range of bifurcation angles in Figure \ref{fig:w2_scans}B. The first minimum on the scan (at zero degrees) is the water dimer minimum, the second minimum near 120 degrees is when the original free \ce{O-H} forms an h-bond, and the maximum near 240 degrees corresponds to the point when the two oxygen atoms point directly at one another. Throughout the entire scan, the agreement between CMM and EDA is excellent. This type of bifurcation motion is known to be important for tunneling processes in water clusters and is believed to be relevant in ice and liquid water.\cite{brown1998bifurcation,richardson2016concerted}

\begin{figure}[H]
\includegraphics*[width=0.95\textwidth]{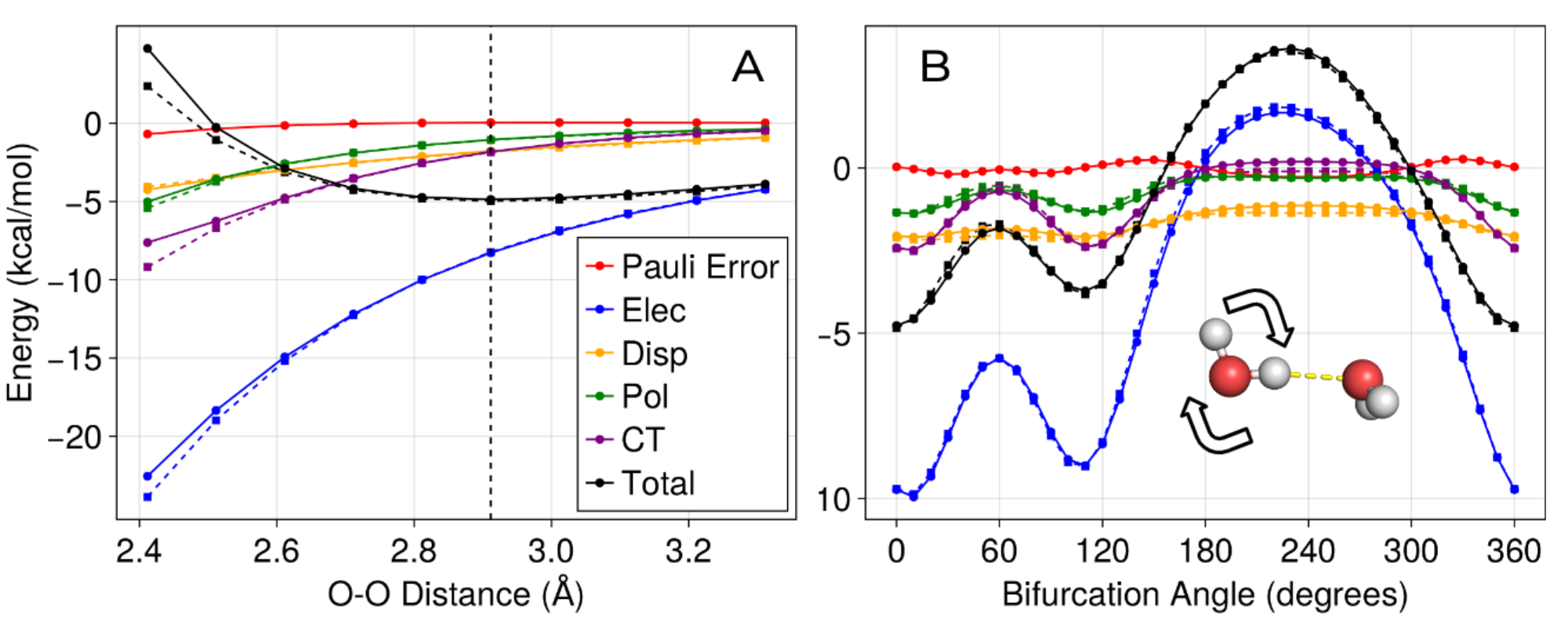}
  \caption{\textit{Water dimer scans with the CMM model (circles, solid line) compared against the  $\omega$B97X-V/def2-QZVPPD DFT reference (squares, dashed).} Rigid scans over two important coordinates on the water dimer potential energy surface. The same legend applies to both figures. Note that we plot the error in Pauli repulsion to keep the y-axis compact. (A) Scan along the oxygen-oxygen distance beginning from the equilibrium water dimer geometry which is indicated by the vertical dashed line. (B) Scan along the bifurcation angle of the water dimer over the full 360 degree rotation. The bifurcation motion is shown with arrows indicating the motion. 
  }
  \label{fig:w2_scans}
\end{figure}
\noindent

In Figure 3, we show a breakdown of the EDA energies into 2-body and 3-body contributions for a scan along the \ce{OOO} distance of the ring water trimer relative to center of the ring. As with the water dimer, the overall agreement is excellent. The 2-body contributions to each term are reproduced extremely well with the largest errors arising in the electrostatic energy at short distances. This is to be expected since at these distances, the energy contribution from multipoles beyond quadrupole become non-negligible. Note, however, that the \ce{O-O} distance in the water trimer is already around 2.8 \AA, which is the same as the average separation in liquid water, so this compressed region is only visited transiently.

The 3-body energies demonstrate that the CMM model of polarization and charge transfer both reproduce many-body energies into the compressed region with very high accuracy. In

\begin{figure}[H]
\begin{subfigure}{0.7\textwidth}
    \includegraphics*[width=\textwidth]{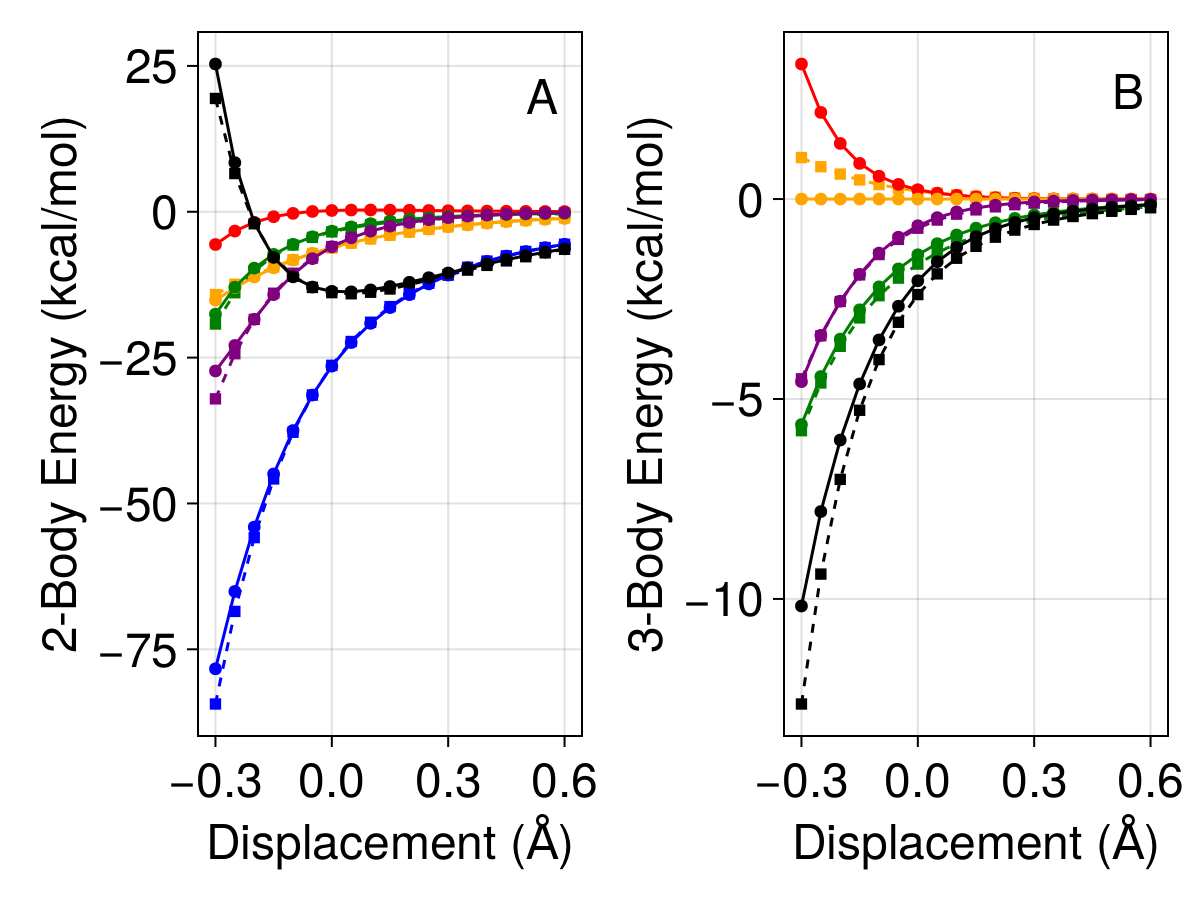}
\end{subfigure}
\begin{subfigure}{0.7\textwidth}
    \includegraphics*[width=\textwidth]{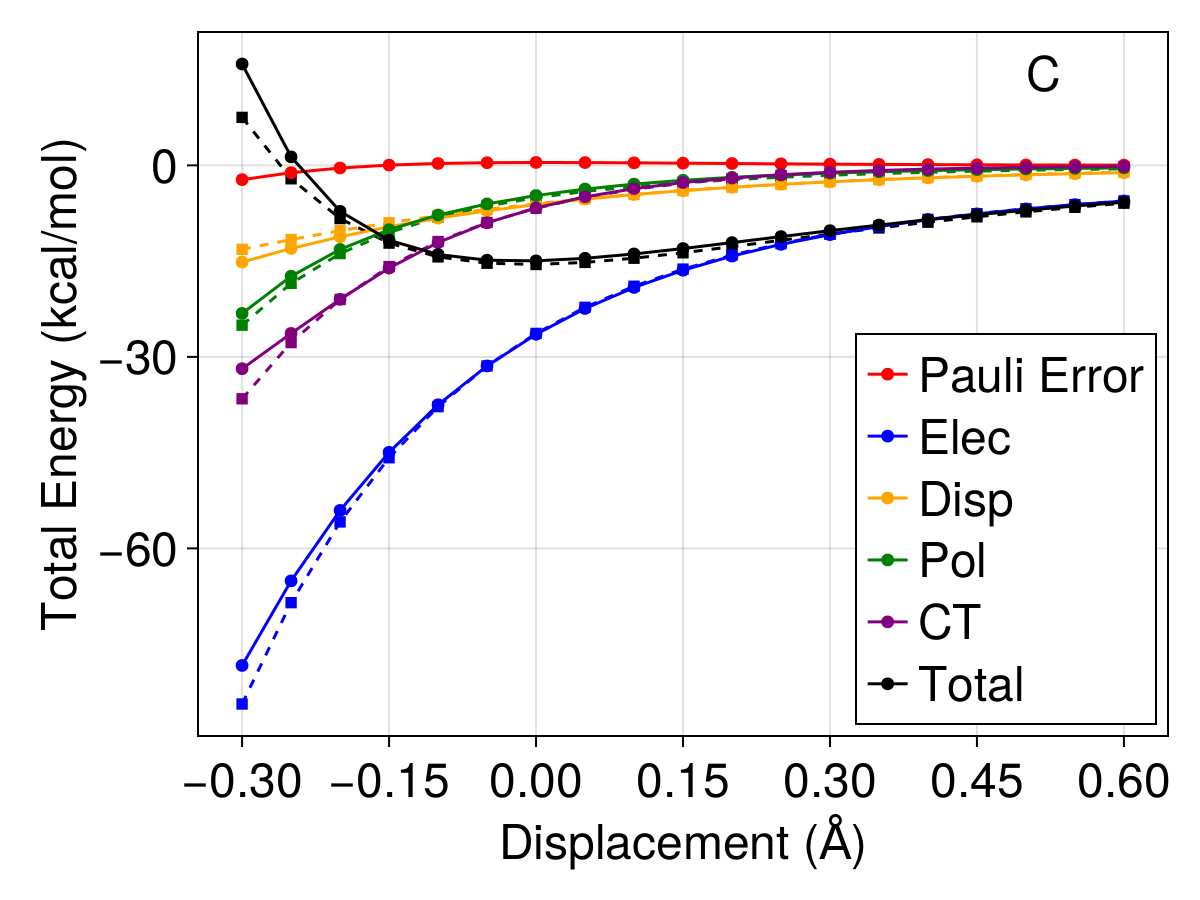}
\end{subfigure}
  \caption{\textit{Water trimer scans with the CMM model (circles, solid line) compared against the $\omega$B97X-V/def2-QZVPPD DFT reference (squares, dashed).}  Rigid scans over the \ce{OOO} distance of the ring water trimer relative to center of the ring. The same legend applies to all figures, and we plot the error in Pauli repulsion to keep the y-axis compact. (A) shows the two-body components of the scan, and (B) shows the same for the 3-body contribution to the energy, and (C) the same for the total contributions to the energy.
  }
  \label{fig:w2_scans}
\end{figure}
\noindent
particular, our model of many-body charge transfer is consistently very accurate even for larger clusters as shown in Table S3. Note, however, that in the compressed region, the total many-body energies are slightly underestimated. This can be seen to arise primarily from the absence of many-body Pauli repulsion in the model.

\begin{figure}[H]
  \includegraphics*[width=0.99\textwidth]{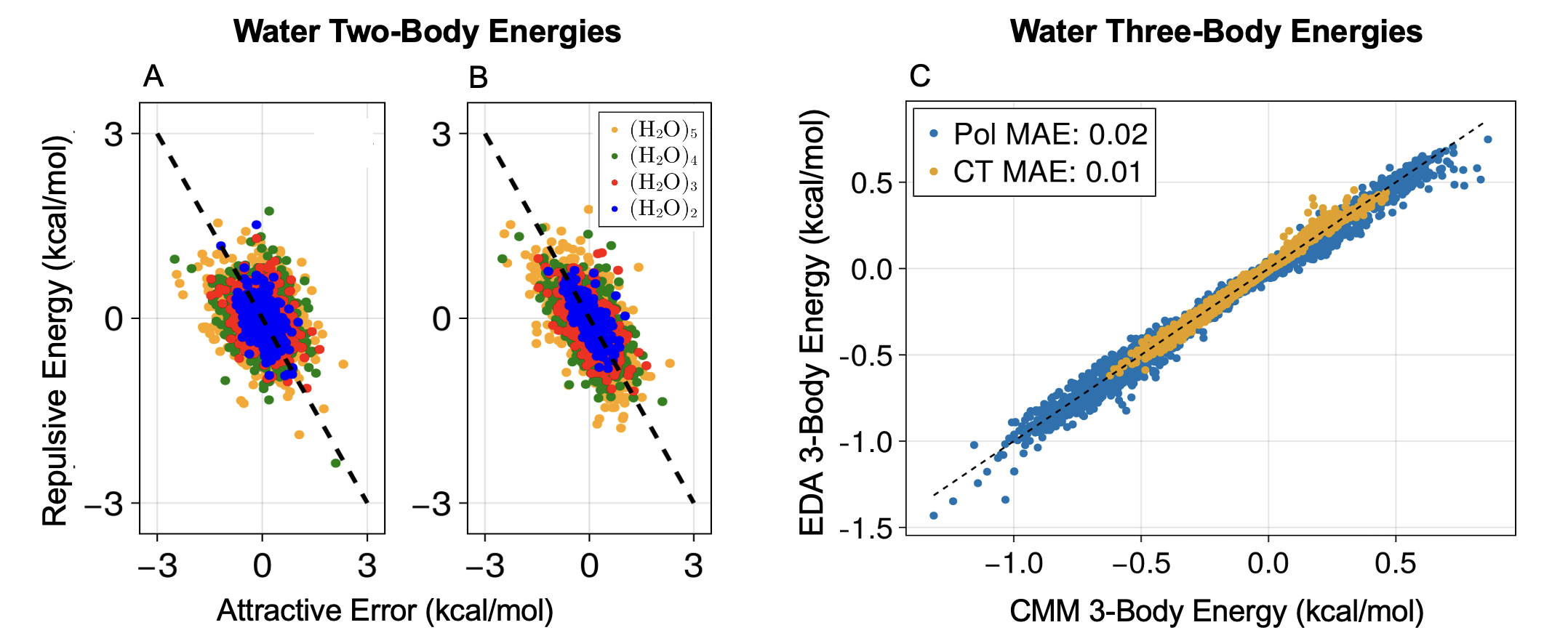}
  \caption{\textit{2-Body and 3-Body Energies of the CMM model.} Correlation of errors in Pauli repulsion against all attractive interactions from EDA including dispersion, electrostatics and CP, polarization, and charge transfer. (A) error correlation without any error fitting, where each EDA term is fit independently. (B) error correlation after allowing the Pauli repulsion to relax against the interaction energy for dimers to improve error cancellation. (C) Correlation plot of the three-body contribution to polarization and
  charge transfer as computed by CMM and with $\omega$B97X-V/def2-QZVPPD.
  All water trimers in this plot span a wide range of configurations, some of which are atypical of liquid water. Since trimers drawn from a cluster may be disconnected, we enforce that the trimer
  have an absolute three-body contribution of at least 0.02 kcal/mol.}
  \label{fig:intermol}
\end{figure}

The CMM model reproduces QM energies and ALMO-EDA energy components for water-water intermolecular interactions with errors no larger than 0.25 kcal/mol (Supplementary Table S2). This can be understood from Figure \ref{fig:intermol} which reports the correlation of errors in all attractive terms against Pauli repulsion both with and without error fitting against total interaction energies. Figure \ref{fig:intermol}A shows that fitting each EDA term independently already correlates attractive and repulsive errors, indicating that some amount of error cancellation is guaranteed, while Figure \ref{fig:intermol}B shows that allowing Pauli repulsion to optimize against the total interaction energy for dimers is only a small correction. 

\noindent

One of the major goals of the CMM is to quantitatively reproduce the many-body contributions to both polarization and charge transfer. To assess how well we have achieved this, we computed the three-body contribution to both polarization and charge transfer with water trimers not used in parameterizing the water model. Figure \ref{fig:intermol}C shows that we obtain excellent agreement with electronic structure for the major three-body contributions to the energy. Indeed, the model manages to capture repulsive and attractive three-body contributions to both polarization and CT despite repulsive three-body contributions to either of those quantities being absent in low-energy water clusters.\cite{heindel2020many} This is important since ion-water clusters, for example, have been shown to have repulsive three-body contributions in many cases.\cite{heindel2021many,herman2021many}

We have also computed the energy of CMM and other advanced water FFs for water clusters for which CCSD(T)/CBS or MP2/CBS benchmark energies are available\cite{herman2023extensive} in Table \ref{tab:benchmark_energies}. The CMM model is seen to do better than MB-Pol and HIPPO, and is nearly as good as q-AQUA. The primary reason we are able to achieve overall excellent total energies on the independent water cluster set is that the individual components of the energy are both accurate and nearly unbiased. While for larger water clusters our model begins to slightly underestimate the energies compared to the CCSD(T)/CBS benchmark references, that is because the DFT method itself, $\omega$B97X-V/def2-QZVPPD, also underestimates the benchmark energies. 

In addition to accurate energetics, many-body FFs should also produce accurate geometries. Table \ref{tab:benchmark_structures} shows the root mean-squared deviation (RMSD) of water cluster structures optimized within each FF and for $\omega$B97X-V/def2-QZVPPD, compared to previously reported structures optimized with either CCSD(T) or MP2.\cite{herman2023extensive} The average RMSD for our model is comparable to MB-Pol and is close to q-AQUA, although the latter difference may be due to the level DFT used in parameterization. Overall, the CMM model can provide accurate geometries at a fraction of the cost of any given electronic structure method.

\begin{table}[H]
  \begin{center}
  \begin{tabular}{llcccccc}
      \hline
      \ce{(H2O)_n}& Isomer & q-AQUA & MB-Pol & CMM & $\omega$B97X-V & HIPPO & Ref. \\\hline
      \ce{(H2O)_2} &  & -4.97 & -4.96                          & -4.93 & -5.00 & -4.96  & -4.99 \\
      \ce{(H2O)_3} & & -15.73 & -15.69                           & -15.03 & -15.77 & -15.77 & -15.77  \\
      \ce{(H2O)_4} &  & -27.35 & -27.12                        & -26.91 & -27.75 &	-26.69  &	-27.39 \\
      \ce{(H2O)_5} &  & -35.71 & -35.94                        & -35.63 & -36.51 &	-34.58  &-35.9 \\
      \ce{(H2O)_6} & Prism & -46.21 & -45.87                   & -45.83 & -46.53 &-46.15 &	-46.2 \\
      \ce{(H2O)_6} & Cage & -45.94 & -45.51                    & -45.29 & -46.30 &	-45.39  &	-45.9 \\
      \ce{(H2O)_6} & Book & -45.21 & -45.19                    & -45.10 & -45.95 &	-44.25  &	-45.4 \\
      \ce{(H2O)_6} & Ring & -43.71 & -44.70                    & -44.18 & -45.07 &	-42.54  &	-44.3 \\
      \ce{(H2O)_7} &  & -57.71 & -57.37                        & -57.20 & -58.08 & -  & -57.4 \\
      \ce{(H2O)_8} & $D_2d$ & -73.32 & -72.28                  & -71.97 & -73.58 & -71.55  & -73.0 \\
      \ce{(H2O)_8} & $S_4$ & -72.93 & -72.35                   & -72.22 & -73.55 & -71.56 & -72.9 \\
      \ce{(H2O)_9} & $D_2dDD$ & -82.87 & -81.67                & -81.49 & -83.00 & -  & -83.0 \\
      \ce{(H2O)_{10}} &  & -94.72 &	-93.07                     & -93.02 &	-94.50 & -  &	-94.6 \\
      \ce{(H2O)_{11}} & 43'4 & -104.23 & -102.17               & -102.09 & -103.77 & -100.23 & -104.6 \\
      \ce{(H2O)_{16}} & Antiboat & -164.87 & -162.20           & -163.38 & -164.20 & -159.63  & -164.6 \\
      \ce{(H2O)_{16}} & 4444-a & -163.10 & -162.98             & -163.20 & -164.28 & -161.84  & -164.2 \\
      \ce{(H2O)_{16}} & 4444-b & -162.54 & -162.87             & -163.15 & -163.84 & -161.56  & -164.1 \\
      \ce{(H2O)_{16}} & Boat a & -164.53 & -161.92             & -162.89 & -164.51 & -159.36  & -164.4 \\
      \ce{(H2O)_{16}} & Boat b & -164.31 & -162.04             & -163.37 & -164.35 & -159.43  & -164.2 \\
      \ce{(H2O)_{17}} & Sphere & -177.56 & -174.15             & -175.45 & -175.78 & -170.68  & -175.7 \\
      \ce{(H2O)_{20}} & ES Prism & -212.49 & -210.20           & -212.09 & -211.98 & -  & -214.2 \\
      \ce{(H2O)_{20}} & FS Prism & -210.63 & -208.46           & -209.22 & -210.12 & - & -211.9 \\
      \ce{(H2O)_{20}} & Fused Cubes & -208.07 & -208.56        & -208.90 & -209.90 & - & -210.6 \\
      \ce{(H2O)_{20}} & Pentag. Dodec. & -199.79 & -197.99     & -198.14 & -201.22 & - & -200.8 \\
      \ce{(H2O)_{25}} & Isomer 2 & -276.50 & -266.04           & -271.37 & -272.02 & - & -276.3 \\\hline
      MAE/n & & \textbf{0.040} &	\textbf{0.156} &	\textbf{0.100} &	\textbf{0.051} & \textbf{0.194}  & - \\\hline
  \end{tabular}
  \end{center}
  \vspace{-3mm}
  \caption{\textit{Comparison of various advanced force fields and DFT against benchmark cluster energies.} The reference energies are CCSD(T)/CBS and MP2/CBS values.\cite{herman2023extensive} We use the FF optimized geometries to evaluate energies, and any FF energies which could not be found in the literature are left blank. The bottom row shows the MAE per molecule for all available energies. $\omega$B97X-V/def2-QZVPPD is the reference data method for CMM.  
  }
  \label{tab:benchmark_energies}
\end{table}
\noindent

\begin{table}[H]
  \begin{center}
  \begin{tabular}{llcccc}
      \multicolumn{6}{c}{Comparison of Methods on Benchmark Water Cluster Structures} \\\hline
      \ce{(H2O)_n} & Isomer & q-AQUA & MB-Pol & CMM & $\omega$B97X-V \\\hline
      \ce{(H2O)_2} &  & 0.005 &	0.008                    & 0.011 &	0.005 \\
      \ce{(H2O)_3} &  & 0.010 &	0.014                    & 0.018 &	0.008 \\
      \ce{(H2O)_4} &  & 0.008 &	0.024                    & 0.010 &	0.006 \\
      \ce{(H2O)_5} &  & 0.013 &	0.059                    & 0.025 &	0.008 \\
      \ce{(H2O)_6} & Prism & 0.010	& 0.035	             & 0.023	& 0.009	\\
      \ce{(H2O)_6} & Cage & 0.013	& 0.027	                 & 0.025	& 0.018	\\
      \ce{(H2O)_6} & Book & 0.010	& 0.029	                 & 0.061	& 0.009	\\
      \ce{(H2O)_6} & Ring & 0.013	& 0.043	                 & 0.014	& 0.010	\\
      \ce{(H2O)_7} &  & 0.016 &	0.041                    & 0.046 &	0.025 \\
      \ce{(H2O)_8} & $D_2d$ & 0.006	& 0.041              & 0.018	& 0.004	\\
      \ce{(H2O)_8} & $S_4$ & 0.007	& 0.019              & 0.017	& 0.005	\\
      \ce{(H2O)_9} & $D_2dDD$ & 0.089 &	0.116            & 0.044 &	0.052 \\
      \ce{(H2O)_{10}} &  & 0.012 & 0.049                   & 0.025 & 0.010  \\
      \ce{(H2O)_{11}} & 43'4 & 0.034 & 0.065               & 0.024 & 0.017  \\
      \ce{(H2O)_{16}} & Antiboat & 0.023 & 0.064           & 0.032 & 0.017  \\
      \ce{(H2O)_{16}} & 4444-a & 0.039 & 0.038             & 0.034 & 0.015  \\
      \ce{(H2O)_{16}} & 4444-b & 0.040 & 0.049             & 0.031 & 0.029  \\
      \ce{(H2O)_{16}} & Boat a & 0.023 & 0.038             & 0.032 & 0.016  \\
      \ce{(H2O)_{16}} & Boat b & 0.028 & 0.057             & 0.060 & 0.016  \\
      \ce{(H2O)_{17}} & Sphere & 0.039 & 0.063             & 0.039 & 0.022  \\
      \ce{(H2O)_{20}} & ES Prism & 0.042 & 0.056           & 0.056 & 0.024  \\
      \ce{(H2O)_{20}} & FS Prism & 0.047 & 0.050           & 0.033 & 0.023  \\
      \ce{(H2O)_{20}} & Fused Cubes & 0.067 & 0.050        & 0.034 & 0.029  \\
      \ce{(H2O)_{20}} & Pentag. Dodec. & 0.034 & 0.066     & 0.047 & 0.018  \\
      \ce{(H2O)_{25}} & Isomer 2 & 0.029 & 0.049           & 0.054 & 0.023  \\\hline
      RMSD (\AA) & & \textbf{0.026} &	\textbf{0.046} &	\textbf{0.032} &	\textbf{0.017}  \\\hline
  \end{tabular}
  \end{center}
  \vspace{-3mm}
  \caption{Comparison of various advanced force fields  against DFT reference and benchmark cluster structures.\cite{herman2023extensive}The reference structures are optimized at either CCSD(T)/aug-cc-pVDZ or MP2/aug-cc-pVTZ. See original paper for further
  details on structures.\cite{herman2023extensive} We also include $\omega$B97X-V/def2-QZVPPD since this is the reference method for CMM. 
  The bottom row shows the root mean-squared deviation (RMSD) in angstrom for all available structures.
  }
  \label{tab:benchmark_structures}
\end{table}


\section*{Discussion and Conclusions}
This work describes a new approach to modeling pairwise and many-body energies, the Completely Multipolar Model, that formulates all terms of an energy decomposition analysis using the electrical multipoles modulated by rank dependent damping functions that control long-ranged and short-ranged asymptotics. The multipoles introduce anisotropy for Pauli repulsion, help formulate fluctuating charges and inducible dipoles for many-body polarization, and control explicit forward and backward charge flow, which in turn feeds into the polarization model to capture many-body charge transfer. It is important to emphasize that the new physics of CT and polarization will minimize the need to rely on other EDA terms to realize cancellation of errors. The fact that the primary source of error in the description of many-body energies is now in Pauli repulsion and dispersion is a nice illustration of the success of this model. We can now quantitatively model many-body contributions to both polarization and charge transfer which, for hydrogen-bonded systems, are likely to always be the dominant source of non-additivities. While dispersion is quite small in water, it is a very important term in many nonpolar systems where dispersion is the dominant interaction. Many-body dispersion can be easily incorporated into the CMM by including a coupling between the dispersion coefficients and the electric field, just like other intermolecular terms and their underlying connection EDA.

One of the key aspects of the CMM is the recognition that there is a common functional form for all EDA terms that describe damping functions based on electrical multipoles as opposed to the density overlap hypothesis developed in other advanced FFs. The density overlap model proposes that short-range interactions are proportional to the overlap of a model density, $V_{sr}=a_ia_jP(b_{ij}r_{ij})e^{-b_{ij}r_{ij}}$, such as the Slater density\cite{van2016beyond,van2018new,rackers2021polarizable}. One of the difficulties associated with the density overlap model, however, is how to include anisotropy in the interactions. Van Vleet et al. have shown that one can include anisotropy by expanding the atom-specific proportionality constants, $a_i$, in terms of spherical harmonics.\cite{van2018new} 
The CMM is markedly different by allowing not just permanent electrostatics and polarization to rely on electrical multipoles, but all EDA terms to share anisotropy through the electric multipoles to encode the size and shape of each atom. This transferability is because the anisotropies of Pauli repulsion and charge transfer are in fact strongly correlated to the electrical anisotropy.

The most important damping parameters for controlling the short-range interactions are the exponential range parameters, $b_i$. One interpretation of this parameter is as the atomic size, which would seem to imply that the parameter should be identical regardless of interaction type. In our model the range for exponential interactions are not strictly identical for each EDA term due to exchange coupling. This is because any interaction necessarily modifies the system wavefunction, and since the wavefunction must be antisymmetric, all terms will couple to exchange. Since the interactions themselves have different ranges and scaling, it seems sensible that the onset of coupling to exchange will also occur at different spatial ranges for each interaction. The best insights we are aware of for this observation comes from the calculations of Tang and Toennies, specifically for exchange-dispersion coupling.\cite{tang1992damping} They show that for the quantum drude oscillator, under the perturbation of an exchange potential, the value of $b_i$ becomes a function of $r$ given by $b(r)=-\frac{d}{dr}\ln{V^{Pauli}(r)}$. The effect of this modification is to cause the damping of dispersion for the quantum drude oscillator to be mediated by a range parameter that is operative at longer ranges than exchange itself. As far as we are aware, there are no analytic results describing if the same modification of the range parameter applies to terms other than dispersion, and further theoretical work is required to address this question. 

When the complete non-bonded CMM model is combined with the one-body potential, we also ensure that the force field reproduces all physically relevant monomer properties including the dipole moment, dipole derivatives, molecular polarizability, and polarizability derivatives for water. Furthermore the CMM model can reproduce the structure-frequency correspondence central to hydrogen-bonded vibrations using a field-dependent contribution to the bonding potential. The necessity of coupling the bonding potential to the environment to accurately reproduce structure-frequency relationships has been shown to reproduce the O-H signatures in a recent Raman theory for water\cite{lacour2023predicting}, and has been applied to a force field for the first time here. We also show that fluctuating charges greatly improve the accuracy of polarizability derivatives, which are essential for computing Raman spectra, which polarizable force fields have historically modeled very poorly. This approach is easily extensible to other force fields and should immediately improve spectroscopic predictions.

In our view the ability to systematically improve a FF is inextricably connected to using EDA as the source of data.\cite{Mao:2021:EDA-review} There is too much noise in the total energy and forces for some of the parameters used here to be fit effectively without EDA. Therefore, the power of EDA for model development is that it breaks down a quantum mechanical energy into terms with well-defined distance-dependent scaling. Not only that, but by allowing for the analysis of forces on a term-by-term basis using FDA,\cite{aldossary2023force} deficiencies in the model are easy to track down. Indeed, we suspect that field-dependent bonding potentials have not (to our knowledge) been used in force fields specifically because their importance is only apparent when the electrostatic forces are separated from the rest of the forces. 

In summary, the CMM can be thought of as a framework enabling systematically improvable descriptions of intermolecular energies and forces that addresses four primary issues. First, we wanted to guarantee that the model respects both long-range and short-range asymptotes since many force fields have failed to accurately account for the overlaping region. Second, we aimed to reproduce as many monomeric properties relevant to intermolecular forces as possible. Third, we desired to maintain a rigorous separation between intermolecular interactions and intramolecular interactions. Finally, the CMM is designed to be transferable, which we will extend to ions and proteins in future work.

\begin{acknowledgement}
We acknowledge support from the U.S. National Science Foundation through Grant No. CHE-2313791. Computational resources were provided by the National Energy Research Scientific Computing Center (NERSC), a U.S. Department of Energy Office of Science User Facility operated under Contract DE-AC02-05CH11231. S.S. thanks the Dutch Research Council (NWO) Rubicon grant (019.212EN.004) for fellowship support. 
\end{acknowledgement}

\begin{suppinfo}
All functional forms of the damping functions, data details, and fitting procedures are provided.
\end{suppinfo}


\section{Supplementary Notes}
\noindent
\textbf{Damping functions for multipole elctrostatics and charge penetration} 
Rackers and Ponder have derived the relevant one-center and two-center damping functions for the Slater density.\cite{rackers2021polarizable} The one-center damping functions modify the potential due to the charge, which after solving Poisson's equation yields,
\begin{equation}
    V_i(r_{ij})=\frac{Q_i}{r_{ij}}\left(1-(1+\frac{1}{2}b_ir_{ij})e^{-b_ir_{ij}}\right)
    \label{eq:potential_1}
\end{equation}
The density-density interaction was computed using integral tables from Coulson\cite{coulson1942two},
\begin{equation}
    V_{ij}(r_{ij})=\frac{1}{2}\left(\int V_i(\bm{r_{ij}})\rho_j(\bm{r_{ij}}) d\bm{r}_id\bm{r}_j + \int V_j(\bm{r_{ij}})\rho_i(\bm{r_{ij}}) d\bm{r}_jd\bm{r}_i\right)=\frac{Q_iQ_j}{r_{ij}}f_{ij}^{overlap}(r_{ij})
    \label{eq:potential_2}
\end{equation}
\noindent
The core-shell interactions are damped according to
\begin{equation}
  \bm{T}_{ij}^{damp}=
  \begin{bmatrix}
    1 & \nabla & \nabla^2 \\
  \end{bmatrix}\cdot
  \left(\frac{1}{r_{ij}}f_{ij}^{damp}(r_{ij})\right)
  \label{eq:T_damp}
\end{equation}
\noindent
while the corresponding interaction tensor for shell-shell damping is written as:

\begin{equation}
  \bm{T}_{ij}^{overlap}=
  \begin{bmatrix}
    1 & \nabla & \nabla^2 \\
    \nabla & \nabla^2 & \nabla^3 \\
    \nabla^2 & \nabla^3 & \nabla^4 \\
  \end{bmatrix}\cdot
  \left(\frac{1}{r_{ij}}f_{ij}^{overlap}(r_{ij})\right)
  \label{eq:T_overlap}
\end{equation}
\noindent
The damping functions $f_{ij}^{damp}(r_{ij})$ and $f_{ij}^{overlap}(r_{ij})$ themselves take the following forms.
\begin{subequations}
  \begin{equation}
    f_{ij}^{damp}(r_{ij})=1-\left(1+\frac12b_{j}r_{ij}\right)e^{-b_{j}r_{ij}}
    \label{eq:damp_a}
  \end{equation}
  \begin{equation}
    f_{ij}^{overlap}(r_{ij})=1-\left(1+\frac{11}{16}b_{ij}r_{ij}+\frac{3}{16}(b_{ij}r_{ij})^2+\frac{1}{48}(b_{ij}r_{ij})^3\right)e^{-b_{ij}r_{ij}} 
    \label{eq:damp_b}
  \end{equation}
\end{subequations}
\noindent
The damping function in Eq. \ref{eq:damp_a} can be derived directly from the form
of the Slater density by computing its electrostatic potential.
The damping function in Eq. \ref{eq:damp_b} can be derived from a symmetrized coulomb integral where each
density interacts with the damped potential generated by the other density.\cite{rackers2021polarizable}
Finally, it is important to note that these damping functions are the ones which apply
to charge-charge interactions and that as higher-order multipoles are considered, new damping
functions are generated alongside the gradients of $1/r_{ij}$.

The damping functions in this work all naturally take the form of a polynomial times an exponential, $f_k(u)=P_k(u)e^{-u}$ where $u=b_{ij}r_{ij}$. To be consistent with past work, damping functions are reported where the value of $k$ is an odd number indicating the power in the denominator of the relevant interaction tensor. For instance, the dipole-dipole interaction would be written as $f_5\frac{3r_{\alpha\beta}}{r^5}-f_3\frac{\delta_{\alpha\beta}}{r^3}$. Since the various $f_k$ can always be written as a polynomial times an exponential, we report only the coefficients of the polynomial, $P_k$, as this is the only unique information. That is, we report the polynomials in a vector representation, $P_k=[c_0, c_1, \cdots, c_n]$ with the understanding that the basis functions are monomials, $[u^0, u^1, \cdots, u^n]$.\\

\noindent
\textit{One-Center Damping Functions:}
\begin{align*}
    P_1^{damp}&=\left[1, \frac12\right] \\
    P_3^{damp}&=\left[1, 1, \frac12\right] \\ 
    P_5^{damp}&=\left[1, 1, \frac12, \frac16\right] \\
    P_7^{damp}&=\left[1, 1, \frac12, \frac16, \frac{1}{30}\right] \\
    P_9^{damp}&=\left[1, 1, \frac12, \frac16, \frac{4}{105}, \frac{1}{210}\right] \\
    P_{11}^{damp}&=\left[1, 1, \frac12, \frac16, \frac{5}{126}, \frac{2}{315}, \frac{1}{1890}\right]
\end{align*}

\noindent
\textit{Two-Center Damping Functions:}
\begin{align*}
    P_1^{overlap}&=\left[1, \frac{11}{16}, \frac{3}{16}, \frac{1}{48}\right] \\
    P_3^{overlap}&=\left[1, 1, \frac{1}{2}, \frac{7}{48}, \frac{1}{48}\right] \\ 
    P_5^{overlap}&=\left[1, 1, \frac{1}{2}, \frac{1}{6}, \frac{1}{24}, \frac{1}{144}\right] \\
    P_7^{overlap}&=\left[1, 1, \frac{1}{2}, \frac{1}{6}, \frac{1}{24}, \frac{1}{120}, \frac{1}{720}\right] \\
    P_9^{overlap}&=\left[1, 1, \frac{1}{2}, \frac{1}{6}, \frac{1}{24}, \frac{1}{120}, \frac{1}{720}, \frac{1}{5040}\right] \\
    P_{11}^{overlap}&=\left[1, 1, \frac{1}{2}, \frac{1}{6}, \frac{1}{24}, \frac{1}{120}, \frac{1}{720}, \frac{1}{5040}, \frac{1}{45360}\right]
\end{align*}

\noindent
\textit{Mutual Polarization Damping Functions:}
\begin{align*}
    P_1^{pol}&=\left[1, \frac{3}{4}, \frac{1}{2}, \frac{1}{32}, \frac{1}{64}\right] \\
    P_3^{pol}&=\left[1, 1, \frac{1}{4}, \frac{7}{16}, -\frac{1}{64}, \frac{1}{64}\right] \\ 
    P_5^{pol}&=\left[1, 1, \frac{5}{12}, \frac{1}{12}, \frac{29}{192}, -\frac{1}{64}, \frac{1}{192}\right]
\end{align*}

\noindent
\textit{Finding new Polynomial Coefficients:} It can be shown that the coefficients of the higher-order polynomials can be derived from the lower-order coefficients in a simple form. Namely, given an initial damping function, $f_1=P_{1}(u)e^{-u}$, the coefficients for $P_3$ and $P_5$ can be expressed in terms of the lower-order polynomials and their $n$th order derivatives, $P_k^{(n)}$,
\begin{align*}
    P_3&=P_1 -(P_1^{(1)}-P_1)u \\
    P_5&=P_3 -\frac{1}{3}(2P_1^{(1)}-P_1^{(2)}-P_1)u^2
\end{align*}

\vspace{3mm}

\noindent
\textbf{Reference Data and Parameterization Procedure.}
Our model is parameterized using water clusters of size \ce{(H_2O)_n} with n=2-5. We use 2400 dimers, trimers, tetramers, and pentamers extracted from various minimized cluster geometries. We additionally generated 4800 pseudo-random water dimers based on a Sobol sequence. We follow exactly the same procedure as described elsewhere.\cite{misquitta2008first} All sampled clusters are available with the paper. All energies and ALMO-EDA calculations used in fitting parameters of the force field are computed at the $\omega$B97X-V level of DFT theory\cite{Mardirossian2014}, and using the def2-QZVPPD basis set\cite{rappoport2010property}, and using the Q-Chem software package\cite{Epifanovsky2021}. All distributed multipole calculations were carried out in the Orient program.\cite{stone2002orient} 

We fit each term against only the EDA contribution to that particular energy component. Optimization of parameters is done using simple gradient descent against the root mean-square deviation (RMSD) of predicted and EDA energies. For electrostatics and Pauli repulsion, we only use dimers in the fitting process since electrostatics is strictly pairwise-additive and Pauli repulsion is nearly so. For these terms, 200 random water dimers from the datasets described above are used in fitting whereas for other many-body terms we use 200 random water dimers, trimers, tetramers, and pentamers from the datasets described above.

When parameterizing electrostatics, we optimize against two objectives. First, we ensure that the dipole derivatives at the equilibrium geometry of water are correct (this can be achieved nearly exactly). Second, we optimize against the distributed multipole electrostatic energy described in the main text.  We then freeze the total charges and dipoles on each atom so that the dipole derivatives will remain correct. Next, we fit the value of the core charges, $Z$ and electrostatic exponents, $b_{elec}$, on each atom with respect to the total electrostatic energy from EDA. We also allow the quadrupoles to relax against the total electrostatic energy as a form of compensation for the lack of higher-order multipoles.

The Pauli repulsion term is first fit against the RMSD of the corresponding EDA energy. At the final step of the parameterization, the repulsion parameters are then allowed to relax against the total interaction energy and interaction forces for only dimers. Using the forces is essential to get meaningful values of the charge flux parameter used in Pauli repulsion. This procedure essentially results in improved error cancellation which we find greatly improves the robustnes of the force field. Since only the Pauli repulsion parameters are modified, this is not expected to inhibit transferability to other systems. It should be noted that we only allow the Pauli repulsion to optimize against dimers so that it cannot correct errors in the many-body contributions. 

The charge transfer energy and dispersion energies are simply fit against the RMSD from their EDA energies. Dispersion has a large enough many-body contribution that if only dimers are used in the fitting, one will systematically over-estimate the dispersion energy since many-body dispersion is usually repulsive. There are methods for modeling many-body dispersion, but we have not included such terms in the current model.\cite{anatole2010two,van2018new} 

\vspace{3mm}

\noindent
\textbf{Environment-dependent bonding potential for the water dimer}
In the course of developing the CMM model, we found that we were unable to reproduce the expected correlation between change in bond length and change in harmonic frequency for \ce{O-H} stretches.\cite{boyer2019beyond} We ultimately determined that the failure of our model to reproduce this correlation was due to a lack of coupling between our bonding potential and the environment. To that end, we extended our Morse potential to be field-dependent in a manner first described elsewhere.\cite{boyer2019beyond} In short, the field-dependence of the potential requires specifying the first and second dipole derivatives along the \ce{O-H} stretch.

We computed the dipole derivatives from electronic structure as follows. First, we scan along an \ce{O-H} stretch and compute the total dipole moment, $\mu$, at each point along the scan. This dipole moment is then projected onto the \ce{O-H} bond vector, $\mu_{\mathrm{OH}}=\frac{R_{\mathrm{OH}}\cdot \mu}{|R_{\mathrm{OH}}|}$. This allows us to isolate how the dipole changes as the \ce{O-H} vibrates in the gas-phase. Note that the large positive second dipole derivative characteristic of hydrogen-bonded water arises from interactions and therefore the appropriate scan is that of an isolated molecule, not one in an environment.

\vspace{3mm}

\noindent
\textbf{Force decomposition analysis}
To summarize the large effect the environment-dependent bonding potential has on the forces within CMM, we  consider a water dimer where each monomer is fixed at the $\omega$B97X-V/def2-QZVPPD geometry in order to eliminate the geometric distortion forces which are not relevant here. We now compute the forces due to each term in EDA using force decomposition analysis\cite{aldossary2023force}, and compare them against the same forces predicted by CMM with and without a field-dependent morse potential in Supplementary Table \ref{tab:dimer_forces}.

The net effect of the field-dependent Morse potential and charge transfer correction is to shift the force from the oxygen atom to the hydrogen atom participating in a hydrogen bond. The improvement in accuracy in electrostatics and polarization forces are particularly notable since this correction involves no free parameters. The improvement in Pauli repulsion is also quite large, correcting about half the force error at negligible computational cost. The charge transfer forces also remove about half the force error along the \ce{O-H} bond. Note that the Pauli repulsion forces can be quite a bit more accurate with the present model, as shown in the parentheses in Supplementary Table \ref{tab:dimer_forces}. Since we use the Pauli repulsion to correlate errors in the model, the Pauli forces compensate for remaining errors in the dispersion and charge transfer forces.

\vspace{3mm}

\noindent
\textbf{Many-body expansion analysis}
It is interesting to see the 2-body and many-body contributions to each component of the energy for whole water clusters. To that end, we computed the 2-body and many-body contributions for a subset of the reference structures used in Tables 3 and 4. A comparison of these quantities as computed with $\omega$B97X-V/def2-QZVPPD and CMM is shown in Supplementary Table \ref{tab:mbe}.  It is evident that the model is generally very accurate at reproducing each term, as should be expected given the excellent performance of the model in reproducing total energies and optimized
structures. 

However, there are a couple of apparent shortcomings worth discussing. First, we currently do not include many-body dispersion which is very small but not entirely negligible, especially considering it is the only term besides Pauli repulsion which is repulsive. While we may explore adding many-body dispersion in the future, it is justifiably neglected for now since many-body charge transfer is generally much more important. In addition, although our model slightly understimates two-body contributions to polarization, the many-body contributions to polarization and charge transfer are both excellent.

\vspace{3mm}

\noindent
\section{Supplementary Figures}

 \begin{figure} [H]
 \centering
    \includegraphics[width=0.99\textwidth]{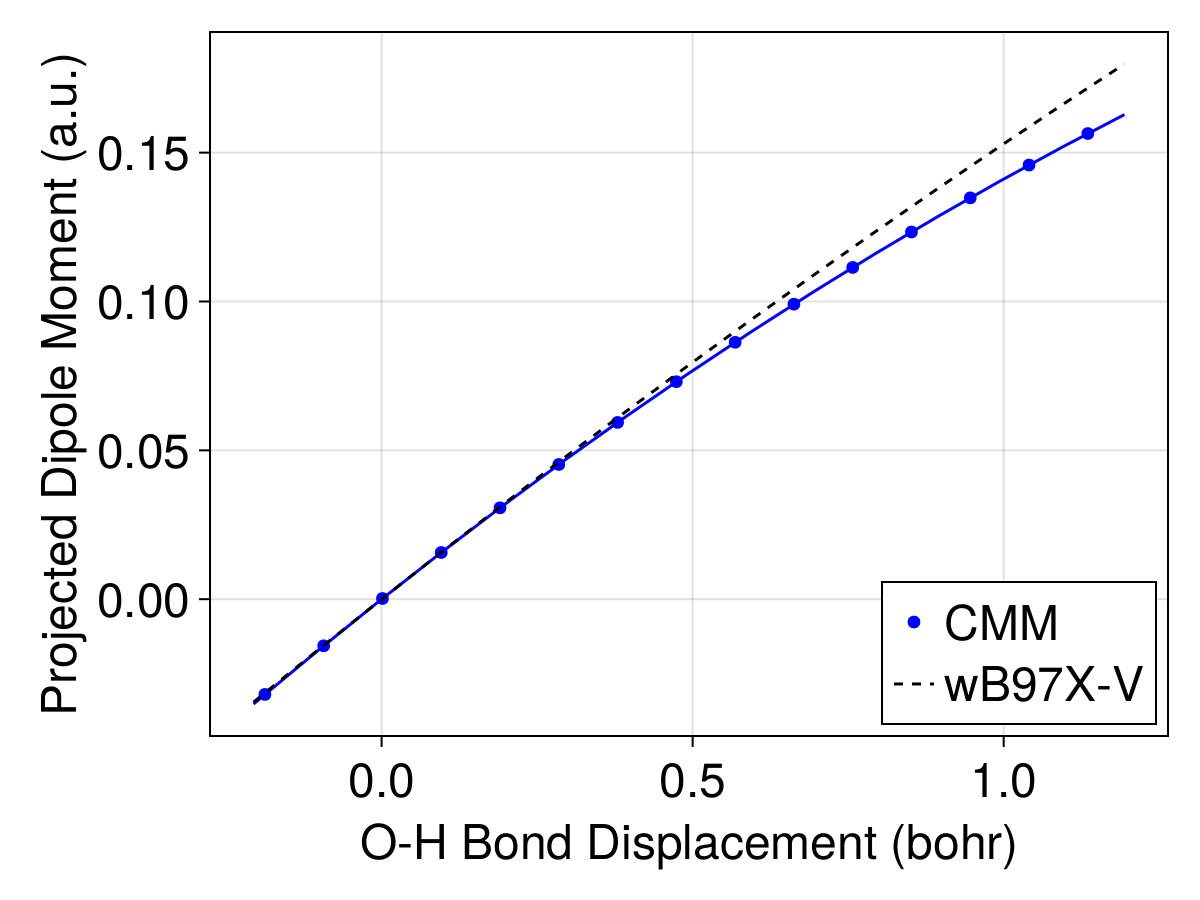}
    \caption{\textit{Projected dipole moments along various O-H stretches.}
    The dipole moment of \ce{(H2O)} is computed with
    $\omega$B97X-V/def2-QZVPPD and CMM as a function of the \ce{O-H} stretch distance. All other degrees of freedom are fixed.
    The dipole moment is projected along the \ce{O-H} stretch unit vector. By fitting a second-order polynomial, we find $\mu_{\mathrm{OH}}^{(1)}=0.165$ $\mu_{\mathrm{OH}}^{(2)}=-0.012$ for $\omega$B97X-V. The CMM dipole derivatives are very similar: $\mu^{(1)}=0.1659$ and $\mu^{(2)}=-0.0248$.
}
    \label{fig:dip_derivatives}
\end{figure}

\vspace{3mm}

\noindent
\section{Supplementary Tables}

\begin{table}[H]
  \begin{center}
  \begin{tabular}{llcccccc}
      \multicolumn{8}{c}{Water Dimer Force Decomposition Analysis Along \ce{O-H} Bond} \\\hline
       Method & Atom & Mod. Pauli & Cls. Elec. & Disp. & Pol. & CT & Total \\\hline
       CMM & \ce{O_{don.}}          & -30.1 & 1.4  & 6.0   & -3.4  & -1.5  & -27.5 \\
       No F.D. Morse & \ce{H_{don.}} & -73.7 & 54.5 & 5.3   & 15.4  & 24.9  & 26.4 \\
       & \ce{O_{acc.}}               & 96.4 & -48.9 & -10.1 & -12.7 & -22.3 & 2.4 \\\hline
       CMM & \ce{O_{don.}} & -15.5 (7.50) & -13.7 & 6.0   & -5.6  & -8.7 & -37.5 \\
       & \ce{H_{don.}}      & -88.5 (-111.9) & 70.0  & 5.3   & 17.7  & 31.9 & 36.4 \\
       & \ce{O_{acc.}}      & 96.4 (95.5)  & -51.7 & -10.1 & -13.0 & -22.0 & -0.4 \\\hline
       $\omega$B97X-V & \ce{O_{don.}} & 3.92 & -15.0  & 0.75  & -8.4  & -15.6 & -34.3 \\
       & \ce{H_{don.}}                & -105.7 & 72.4 & 10.8  & 18.9  & 39.3  & 35.8 \\
       & \ce{O_{acc.}}                & 92.3 & -52.3  & -10.3 & -10.4 & -19.6  & -0.50 \\\hline
  \end{tabular}
  \end{center}
  \caption{\textit{Comparison of the forces projected along the \ce{O-H} bond of a water dimer due to each component in the EDA.} Forces predicted by CMM are computed with and without the field-dependent morse potential described in the main text. The first block of entries is CMM with no field-dependent Morse potential, the second block is CMM, and the third is $\omega$B97X-V/def2-QZVPPD. All forces are in kJ/mol/\AA. \ce{O_{don.}} and \ce{H_{don.}} are the oxygen and hydrogen atoms donating the hydrogen bond and \ce{O_{acc.}} is the oxygen accepting the hydrogen bond. All forces are projected along the \ce{O-H} bond vector. The numbers in parentheses are what the forces before error fitting with Pauli repulsion.}
  \label{tab:dimer_forces}
\end{table}

\begin{table}[H]
  \begin{center}
  \begin{tabular}{lccccc}
      \multicolumn{5}{c}{MAE of Force Field EDA Terms (kcal/mol)} \\\hline
       & \ce{(H2O)2} & \ce{(H2O)3} & \ce{(H2O)4} & \ce{(H2O)5} \\\hline
      Pauli (no error fit)   & 0.134 (1.513)  & 0.209 (0.538) & 0.277 (0.491) & 0.329 (0.400) \\
      Pauli                  & 0.195 (-0.064)  & 0.297 (-0.007) & 0.387 (0.117) & 0.506 (-0.098) \\
      Electrostatics         & 0.123 (-0.268) & 0.206 (0.189) & 0.283 (-0.092) & 0.348 (-0.061) \\
      Dispersion             & 0.069 (0.090)  & 0.092 (-0.124) & 0.109 (-0.407) & 0.149 (-0.420) \\
      Polarization           & 0.047 (-0.243)  & 0.088 (-0.041) & 0.122 (0.437) & 0.155 (0.506) \\
      Charge Transfer        & 0.102 (-0.610)  & 0.159 (-0.207) & 0.218 (-0.278) & 0.264 (-0.197) \\\hline
      Interaction            & 0.089 (1.158) & 0.166 (0.875) & 0.226 (0.399) & 0.290 (0.428) \\\hline
  \end{tabular}
  \end{center}
  \vspace{-3mm}
  \caption{\textit{Comparison of the mean absolute error (MAE) of all terms in the EDA against CMM predictions  for hydrogen-bonded water dimers, trimers, tetramers, and pentamers.} 
  In total, there are 2400 each of dimers, trimers, tetramers, and pentamers.
  The first row shows the Pauli repulsion energy without inclusion of error fitting
  while the second row is the Pauli repulsion used in the final model which is calibrated
  to maximize error correlation. MAE in kcal/mol }
  \label{tab:mae}
\end{table}

\begin{table}
    \begin{tabular}{llcccccccccc}
      \multicolumn{12}{c}{Comparison of Many-Body Expansion for EDA Components (kcal/mol)} \\\hline
      \ce{(H2O)_n}& Component & \multicolumn{2}{c}{Cls. Elec.} & \multicolumn{2}{c}{Mod. Pauli} & \multicolumn{2}{c}{Disp.} & \multicolumn{2}{c}{Pol.} & \multicolumn{2}{c}{CT} \\\hline
      & & CMM & DFT & CMM & DFT & CMM & DFT & CMM & DFT & CMM & DFT \\\hline
      \ce{(H2O)_3} & 2-Body    & -27.02 & -26.77 & 29.51 & 28.87 &  -6.31  & -6.28  &	-3.41  & -3.48  &	-6.29  & -6.08 \\
             & $\ge$ 3-Body    & -      & -      & 0.0   & -0.24 &   0.0   & 0.21   &	-1.42  & -1.63  &	-0.77  & -0.74 \\\hline
                    & Total    & -27.02 & -26.77 & 29.51 & 28.63 &  -6.31  & -6.07  &	-4.84  & -5.11  &	-7.07  & -6.82 \\\hline
      \ce{(H2O)_6} & 2-Body    & -79.01 & -78.43 & 88.93 & 88.40 &	-19.65 & -19.90 &	-10.67 & -10.94 &	-18.32 & -18.41 \\
      Prism & $\ge$ 3-Body     & -      & -      & 0.0   & -0.62 &	 0.0   & 0.83   &	-5.82  & -6.40  &	-2.93  & -2.80 \\\hline
                    & Total    & -79.01 & -78.43 & 88.93 & 87.78 &	-19.65 & -19.07 &	-16.49 & -17.34 &	-21.26 & -21.21 \\\hline
      \ce{(H2O)_6} & 2-Body    & -78.76 & -78.67 & 90.15 & 90.10 &	-19.13 & -19.36 &	-11.17 & -11.25 &	-19.41 & -19.62 \\
      Cage & $\ge$ 3-Body      & -      & -      & 0.0   & -0.66 &	 0.0   & 0.79   &	-5.76  & -6.26  &	-3.15  & -3.07 \\\hline
                    & Total    & -78.76 & -78.67 & 90.15 & 89.44 &	-19.13 & -18.56 &	-16.94 & -17.51 &	-22.57 & -22.69 \\\hline
      \ce{(H2O)_{10}} & 2-Body & -161.5 & -162.4 & 190.4 & 192.2 &	-37.89 & -38.34 &	-24.69 & -24.92 &	-42.62 & -43.27 \\
                & $\ge$ 3-Body & -      & -      & 0.0   & -0.68 &	0.0    & 1.18   &	-13.50 & -14.92 &	-7.67  & -7.42 \\\hline
                & Total        & -161.5 & -162.4 & 190.4 & 191.5 &	-37.89 & -37.16 &	-38.19 & -39.84 &	-50.29 & -50.69 \\\hline
      \ce{(H2O)_{16}} & 2-Body & -276.8 & -277.7 & 327.2 & 329.0 &	-67.23 & -67.61 &	-42.21 & -42.44 &	-72.41 & -73.15 \\
      Antiboat  & $\ge$ 3-Body & -      & -      & 0.0   & -1.06 &	0.0    & 2.17   &	-23.93 & -27.07 &	-12.98 & -12.37 \\\hline
                & Total        & -276.8 & -277.7 & 327.2 & 327.9 &	-67.23 & -65.44 &	-66.14 & -69.51 &	-85.39 & -85.52 \\\hline
      \ce{(H2O)_{20}} & 2-Body & -354.7 & -355.4 & 419.4 & 420.9 &	-89.17 & -89.20 &	-53.76 & -54.26 &	-92.54 & -92.80 \\
      ES Prism  & $\ge$ 3-Body & -      & -      & 0.0   & -1.66 &	0.0    & 3.03   &	-30.53 & -34.85 &	-16.28 & -15.21 \\\hline
                & Total        & -354.7 & -355.4 & 419.4 & 419.3 &	-89.17 & -86.17 &	-84.29 & -89.11 &	-108.8 & -108.0 \\\hline
    \end{tabular}
  \vspace{-3mm}
  \caption{\textit{Comparison of the 2-body, many-body, and total energies as predicted by CMM and as computed with $\omega$B97X-V/def2-QZVPPD.}
  Both calculations are done at the \textit{ab initio} optimized geometries. Names of isomers, when applicable, are written below the
  cluster size.}
  \label{tab:mbe}
\end{table}

\bibliography{references}

\end{document}